\documentclass{aa}  

\usepackage{graphicx}
\usepackage{txfonts}
\usepackage{hyperref}
\usepackage{xcolor}

\definecolor{nncolor}{rgb}{0.0,0.7,0}
\definecolor{fmcolor}{rgb}{0,0,0.7}
\definecolor{jccolor}{rgb}{0.9,0,0.}

\newcommand{\RE}{R_{\rm E}}

\begin{document}

\title{Estimating the number of planets that PLATO can detect}

\author{F.~Matuszewski
 \inst{1}
 \and
 N.~Nettelmann\inst{1}
 \and
 J.~Cabrera\inst{1}
 \and
 A.~B\"orner\inst{2}
 \and
 H.~Rauer\inst{1,3,4}
 }

 \institute{Institute of Planetary Research, German Aerospace Center, Rutherfordstrasse 2, 12489 Berlin, Germany
 \and
 Institute of Optical Sensor Systems, German Aerospace Center, Rutherfordstr.~2, 12489 Berlin, Germany
 \and
 Department of Geological Sciences, Freie Universit\"at Berlin, 12249 Berlin, Germany
 \and
 Center for Astronomy and Astrophysics, Technical University
Berlin, Hardenberstrasse 36, 10623 Berlin, Germany
}


 
  \abstract
   {The PLATO mission is scheduled for launch in 2026. It will monitor more than 245\,000 FGK stars of magnitude 13 or brighter for planet transit events. Among the key scientific goals are the detection of Earth-Sun analogs; the detailed characterization of stars and planets in terms of mass, radius, and ages; the detection of planetary systems with longer orbital periods than are detected in current surveys; and to advance our understanding of planet formation and evolution processes.}
    {This study aims to estimate the number of exoplanets that PLATO can detect as a function of planetary size and period, stellar brightness, and observing strategy options. Deviations from these estimates will be informative of the true occurrence rates of planets, which helps constraining planet formation models.}
   {For this purpose, we developed the Planet Yield for PLATO estimator (PYPE), which adopts a statistical approach. We apply given occurrence rates from planet formation models and from different search and vetting pipelines for the {\it Kepler} data. We estimate the stellar sample to be observed by PLATO using a fraction of the all-sky PLATO stellar input catalog (PIC). PLATO detection efficiencies are calculated under different assumptions that are presented in detail in the text.}
   {
   The results presented here primarily consider the current baseline observing duration of four years. 
   We find that the expected PLATO planet yield increases rapidly over the first year and begins to saturate after two years. A nominal (2+2) four-year mission could yield about several thousand to several tens of thousands of planets, depending on the assumed planet occurrence rates. We estimate a minimum of 500 Earth-size (0.8--1.25 $\RE$) planets, about a dozen of which would reside in a 250-500d period bin around G stars. We find that one-third of the detected planets are around stars bright enough (V $\leq 11$) for RV-follow-up observations. We find that a three-year-long observation followed by 6 two-month short observations (3+1 years) yield roughly twice as many planets as two long observations of two years (2+2 years). The former strategy is dominated by short-period planets, while the latter is more beneficial for detecting earths in the habitable zone.}
   {Of the many sources of uncertainties for the PLATO planet yield, the real occurrence rates matters most. Knowing the latter is crucial for using PLATO observations to constrain planet formation models by comparing their statistical yields.}


   \maketitle
%

\section{Introduction}

Since the discovery of an exoplanet around a main-sequence star \citep{MayorQueloz95} and the first radius measurement of an exoplanet \citep{Charbonneau00}, over 5\,000 planets have been confirmed to exist in extrasolar systems (NASA Exoplanet Archive\footnote{http://exoplanetarchive.ipac.caltech.edu}). Results from the NASA \textit{Kepler} mission show that the number of exoplanets exceeds the number of stars \citep[e.g.,][]{fressin2013}. Accordingly, \citet{Poleski21} inferred a planet-to-star ratio of $\sim 1.4$ for planets at wide (5--15 AU) orbits.

The \textit{Kepler} mission (2009-2013) has been one of the most successful missions, with more than 2\,700 discovered planets and an additional $\sim$ 2\,000 planetary candidates. This mission was designed foremost to determine the occurrence rate of planets of different sizes and orbital distances around different stellar types, in particular, the occurrence rate of Earth-size planets near the habitable zone (HZ) of Sun-like stars \citep{Borucki10,Batalha14}, where \emph{\textup{Sun-like}} here refers to FGK main-sequence stars.

However, the question of the occurrence rate of planets in dependence on their size and orbital distance around FGK stars is not yet settled. 
To some extent, this is due to the uncertainties in the detection efficiency of the telescopes that were used \citep{Bryson21,Poleski21}, and it is due to challenges in the recognition of false positives \citep{fressin2013} or false negatives \citep{hsu2019}. The uncertainty on the stellar radius also matters because it directly reflects on the inferred planet radius.

Nevertheless, Earth-size objects in the HZ of Sun-like stars have been detected. Using stellar radii from the Gaia DR2-catalog and \textit{Kepler} data from the DR25 release \citep{Thompson2018}, \citet{Bryson21} identified four planets in the 0.8-1.3 Earth-radius ($\RE$) range that reside in the respective HZs of their parent GK stars.
The NASA Exoplanet Archive lists seven confirmed or candidate {\it Kepler} objects of interest (KOIs) in this size range that receive an incident flux within 0.15 and 1.0 that of Earth, consistent with the conservative HZ definition of \citet{Kopparapu13}. However, their masses are unknown, and hence we cannot characterize their composition. 
These planets orbit faint stars at magnitude 13 or fainter, rendering a mass measurement and thus a characterization challenging with current instrumentation.

The mission called PLAnetary Transits and Oscillation of stars (PLATO) \citep{rauer2014plato} has been designed to detect and characterize terrestrial planets in the HZ of Sun-like stars as well as other planetary systems. 
These goals will be achieved by high-precision photometry and long monitoring of bright (V $\leq 13$) FGK stars in the search for transit events. The ESA PLATO mission will be able to characterize the planets by complementing the radius measurements with information on mass and age for the bright end (V$\leq 11$). The age of the system can be inferred from astroseismology, while the planet mass can be measured from the ground by the radial velocity method. Moreover, by detecting a large number of planets, the PLATO observations are expected to help constrain planet formation models (see the PLATO definition study report or ''red book''\footnote{https://sci.esa.int/web/plato/-/59252-plato-definition-study-report-red-book}).

It is clear that the number of planets that PLATO can detect correlates with the intrinsic number of planets. Using the occurrence rate estimates of \citet{fressin2013}, a PLATO planet yield (PPY) of 4\,600 to 11\,000 transiting planets, of which 3-280 will lie in the habitable zone, has been estimated (red book).

Subsequent years have seen improvements in planet occurrence rate estimates and in the characterization of the stellar sample to be observed by PLATO.
This indicates that a re-evaluation of the PLATO capabilities and an update of previous planet yield estimates is required.
Previously, \cite{Heller22} revised the estimate of the PPY for Earth-size planets in the HZ of the bright P1 subsample of stars to be observed by PLATO.
In this study, we aim to estimate the PLATO planet yield across a wide range of planet sizes and orbital distances, using improved estimates for the occurrence rate, the knowledge of the stellar sample to be observed by PLATO, and its detection efficiency. 
To capture current uncertainties in the planet occurrence rates, we employed two different approaches using three different sources. On the one hand, we applied predictions from planet formation models \citep{emsenhuber2020a,emsenhuber2020b}. On the other hand, we used empirical estimates from \textit{Kepler} observations \citep{hsu2019,kunimoto2020}. 
We investigated the PPY in dependence on planet size, orbital distance, and possible mission scenarios because the PLATO observing strategy is not yet fixed.

Results from this study can help in the decision making, in particular, regarding the observing duration of single fields. 
We primarily focus on the current baseline observing duration of four years (as constrained by the ESA schedule planning) and the scenarios as addressed in the red book. 
We recall that PLATO is designed for a six-year observing duration and will have consumables for up to eight years of operations. 
To take advantage of these options, mission extensions can be applied for after launch of the mission. 
These scenarios deserve further studies, which are currently ongoing.

The paper is structured as follows. In Section \ref{sec:inputplanets} we describe the three external planet input occurrence rates we used as input and how we converted them into planet population models (\S \ref{sec:pop}). In Section \ref{sec:stars} we describe the input stellar sample, and in Section \ref{sec:DE} we present our detection efficiency model. 
Section \ref{sec:PYPE} presents our combination of input parameters with which we obtained the  PPY as output, using our new code PYPE. Section \ref{sec:results} contains the results for the PPY for the nominal assumptions (\ref{sec:results1}, \ref{sec:results2}), but also for the assumptions of single-transit detections (\ref{sec:results3}) and the LOP1 stellar fields (\ref{sec:results4}). Given the importance of potentially habitable planets but their poorly known occurrence rate,  we make an estimate of the PPY for this class of planets  in Section \ref{sec:discussion}. We also compare our PPY with the current TESS project candidates. Section \ref{sec:summary} summarizes our conclusions.

\section{Input planetary samples} \label{sec:inputplanets}

We applied three different models for the assumed true population of planets in the orbital period--planet radius space. We labeled them NGPPS, HSU19, and KM20, and we describe them in the following three subsections.

\subsection{The NGPPS model}

The NGPPS model is the generation III version of the new generation planetary population synthesis (NGPPS-76) model of \citet{emsenhuber2020a,emsenhuber2020b}. This work is based on previous population synthesis models with theoretical formation models~\citep{alibert2005models,mordasini2009extrasolara,mordasini2009extrasolarb}.

In the generation III model, the formation of planets in a protoplanetary disk is simulated starting from 100 Moon-size embryos that are distributed between the solar-mass star and a semi-major axis $a$ of 40 AU with a uniform probability in log $a$. The evolution of the disk, the luminosity evolution of the star, the growth of the embryos, and the planet-disk interaction are simulated, including a wide range of physical processes such as planetesimal accretion, gas accretion, migration, and N-body interactions between the protoplanets. The long evolution of the planets and the star is followed for up to several billion years, including possible atmospheric escape, planet cooling, and contraction. Moreover, the NGPPS model takes into account various uncertainties in disk properties and treats collisions and planetesimal capture as a probabilistic process. Different simulation runs yield somewhat different outcomes of multiplanet systems. 
By summarizing the outcome of different runs, this diversity can be captured in the resulting planet population.

\begin{figure}[h]
        \includegraphics[width=0.5\textwidth]{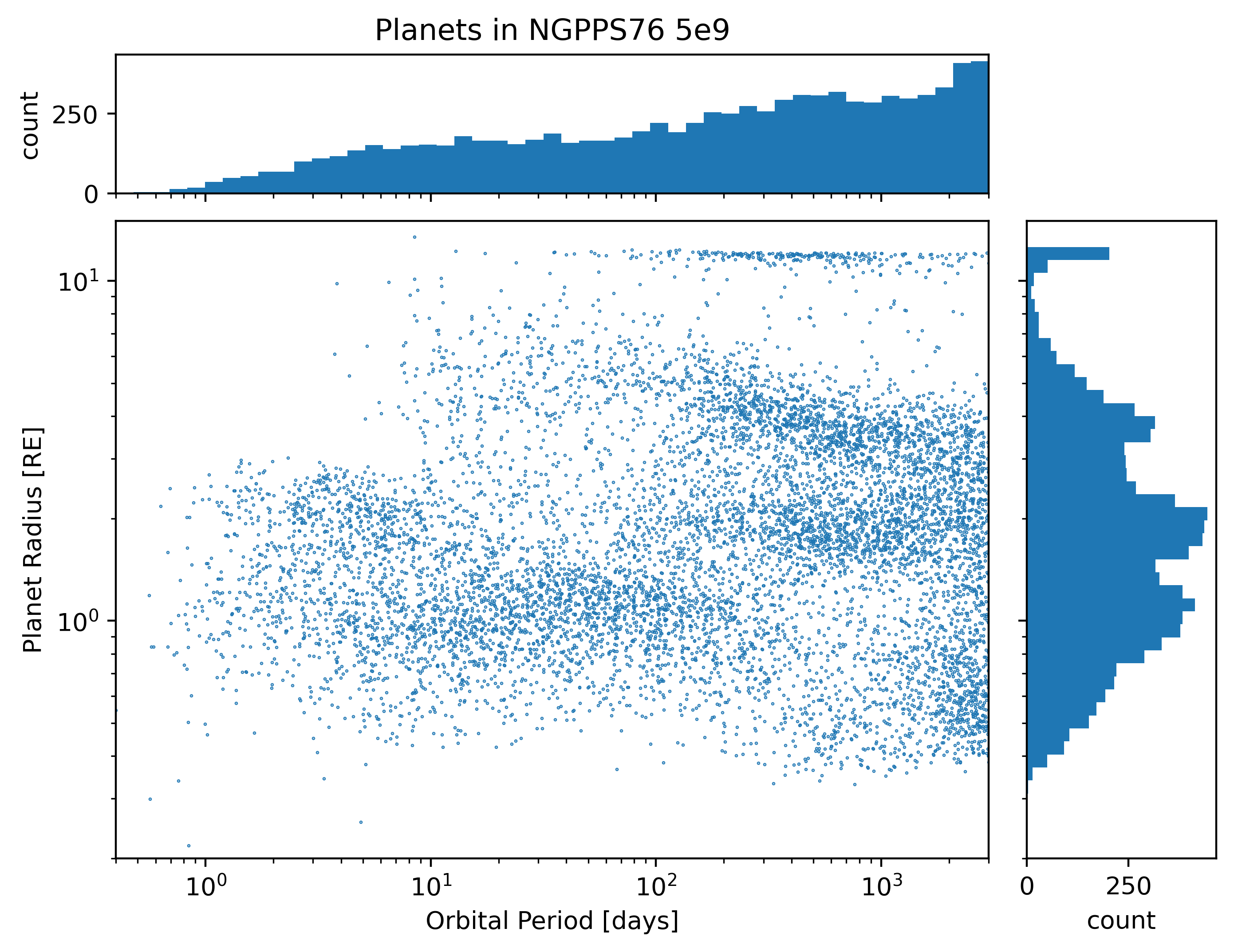}
        \caption{Planets of the NGPPS76 5e9 population model for 1\,000 stars. We show only planets up to an orbital period of eight years, which are of interest for this study. The top panel shows the histogram over orbital period, and the right panel shows the histogram over planet radius.}
        \label{fig:NGPPS76}
\end{figure}

We used the nominal planetary population model NGPPS-76 of \cite{emsenhuber2020b}. It contains the outcome of 1\,000 simulation runs around a 1 $M_{\rm Sun}$ star and starts from 100 embryos. The model was kindly provided to us in form of tables (Ch. Mordasini, pers. comm 2020). We used the particular table for the outcome after 5 Gyr of long-term evolution, as we anticipate that the majority of stars will be mature.
Of the parameters relevant to this work, the table includes the planetary radius $R_p$, the semi-major axis $a$, the orbital inclination, the eccentricity, and a starID for each planet to follow the system to which a planet belongs. A few out of the 1\,000 simulated systems do not harbor any planet. Their starIDs thus do not occur in the table. We calculated the fraction of systems with at least one planet to be $\eta=0.98$ for the NGPPS population model. When we assumed a sample of planets $>$0.5 $\RE$ up to an orbital period of eight years, the average number of planets per star is almost nine. Jupiter-size planets are least abundant, which is consistent with \textit{Kepler} observations \citep{fressin2013,hsu2019,kunimoto2020}, but it is different from the Solar System, where the Neptune-to-Jupiter ratio is 1:1. Moreover, Earth-size planets are formed in the NGPPS model as often as super-Earths and sub-Neptunes, while \textit{Kepler} observations indicate that the latter group is most abundant \citep{Fulton17,kunimoto2020}, although the \textit{Kepler} detection limit for Earth-size planets at orbital periods of months or longer still precludes a definite conclusion \citep{hsu2019}. 
Figure \ref{fig:NGPPS76} illustrates the distribution of planets in this catalog over the radius and orbital period.

\subsection{The Hsu19 planet occurrence rate}

\citet{hsu2019} revisited the empirical occurrence rate of \textit{Kepler} planets around main-sequence FGK stars over a wide range of radii and orbital periods. They used what is still the latest catalog of planets that are characterized  as candidates by the \textit{Kepler} Transiting Planet Search Pipeline (DR25); they used improved stellar parameter values (radius and spectral class)  from the current \emph{Gaia} mission \citep{gaia2018gaia} provided in form of the DR2 catalog. They improved the estimation of the false-negative rate, also known as completeness, of the \textit{Kepler} pipeline in recognizing a planet-caused transit signal as a threshold crossing event, and on its vetting as a planet candidate or false alarm as conducted by the robovetter \citep{Mullally2016,Thompson2018}. 

In particular, \citet{hsu2019} employed a statistical method for the consideration of uncertainties. The inferred planet occurrence rates depend on the probability distributions of imperfectly known parameters. Approximate Bayesian computing  (ABC) is a powerful statistical method particularly in cases when the likelihood functions for the data given the model are poorly known.  \citet{Hsu18} introduced the ABC method to the planet occurrence rate estimation from observational data. With this method, it is possible to predict upper limits in areas of $R_p$--$P_{\rm orb}$ space that are empty because no planet has been detected there yet, which is particularly relevant to Earth-size planets in the HZ of Sun-like stars. Because it can account for observational uncertainties in a forward modeling approach, the ABC method is argued \citep{kunimoto2020} to be more accurate than the inverse detection efficiency method applied in previous works \citep{fressin2013}. Overall, \citet{hsu2019} obtained an average occurrence rate of 2.5 planets per star when areas of upper limits only were excluded. In contrast, the average occurrence rate rises to 5.2 planets per star when these areas are included by using a uniform distribution between zero and the given upper limit.

\subsection{The KM20 planet occurrence rate}

Despite recent improvements, the planet occurrence rate from the \textit{Kepler} data is still not entirely agreed upon.
Uncertainty persists due to our limited knowledge of the host star properties, and thus of the planet radius derived from the transit depth. 
For instance, the \textit{Kepler} pipeline of transiting planet determinations assumes star-averaged radii for the estimation of the \textit{Kepler} detection efficiency. This leaves further room for improvement. 

In an effort to improve the inference of the planet occurrence rates of FGK stars from the \textit{Kepler} data, \citet{kunimoto2020} compiled an independent catalog of \textit{Kepler} planet candidates. They employed the ABC method and included the star-specific radii in the calculation of the detection and vetting efficiencies.
Their analysis is based on the \textit{Kepler} subsample of 116\,637 FGK stars with a radius measurement accuracy of  10\% or better. For these stars, they searched the \textit{Kepler} Q1-Q17 database for transit events. They defined an event as a transit candidate when at least three events with a signal-to-noise ratio S/N>6 were observed. They applied their own vetting procedure to distinguish false positives and false negatives from what otherwise could be considered a planet candidate.
With their method, \citet{kunimoto2020}  achieved a recovery rate of 99\% of the confirmed KOIs and 65\% of the candidate KOIs labeled as such in the NASA exoplanet Archive for the same stellar sample. Their planetary sample comprises 2\,533 planets, which populate 79 out of the $9\times 10$ cells in $P_{\rm orb}$--$R_p$ space between 0.78---400 days and 0.5---16 $\rm \RE$.
Compared to \citet{hsu2019}, their detection and vetting efficiency values are lower for S/N values below 16 by up to a factor of  a few, yielding lower planet occurrence rates where the S/N is this low. Marginalized over all FGK-type stars and all planets, they arrived at an occurrence rate of $1.06^{+0.09}_{-0.07}$ planets per star.
Furthermore, \citet{kunimoto2020} did not find a statistically significant planet radius valley over a large $P_{\rm orb}$ range as \citet{Fulton17} did, but instead found an increased abundance of 2.0-2.8 $\rm R_E$ planets.
The accumulated estimates for $P<50$ and $P<100$ agree well in general with previous works that used the inverse detection efficiency method. We used their occurrence rate diagram, which is marginalized over FGK stars, and call our population created from this input data set KM20.

\subsection{From occurrence rates to population models}\label{sec:pop}

\begin{figure}[t]
    \centering
        \includegraphics[width=0.45\textwidth]{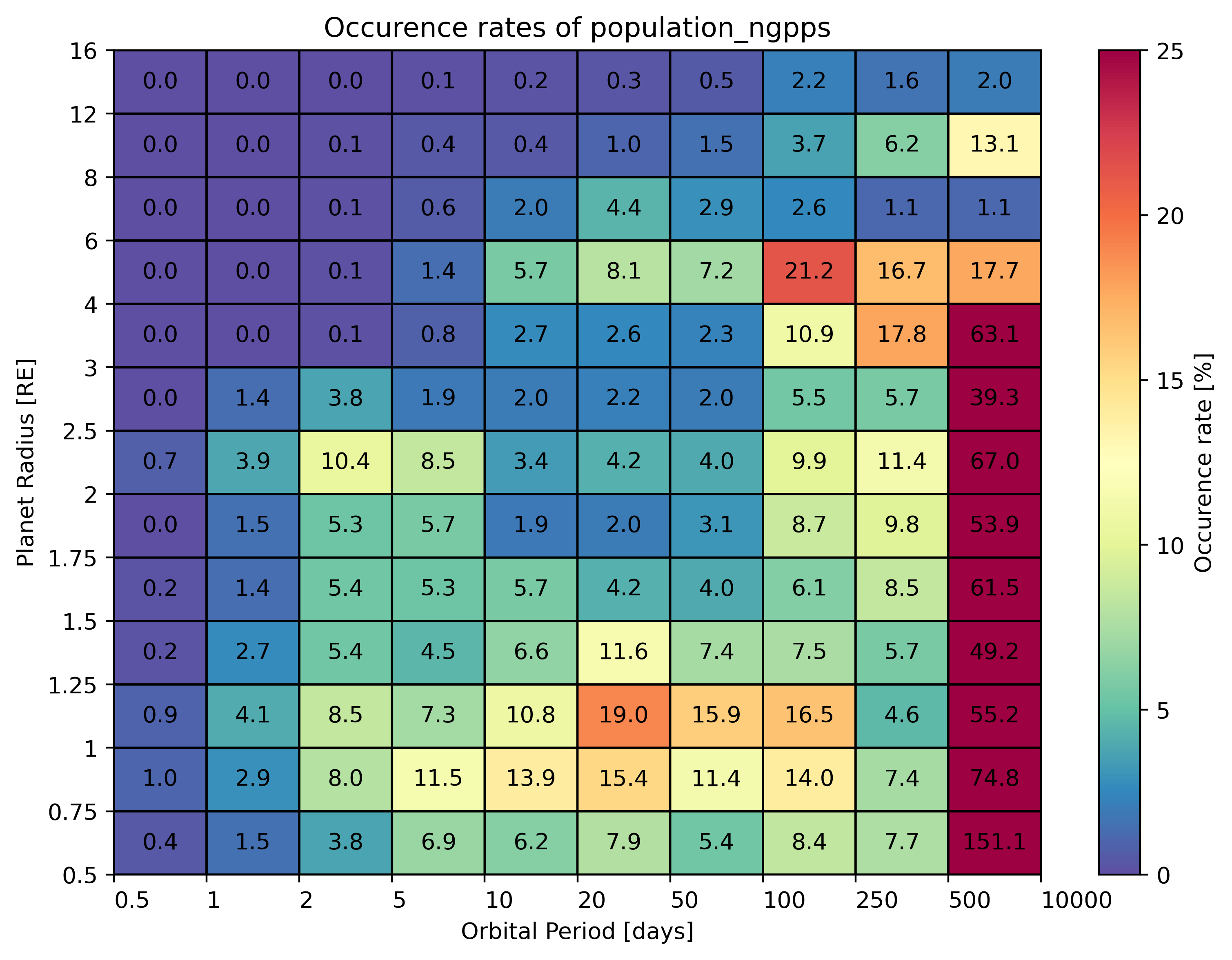}
        \includegraphics[width=0.45\textwidth]{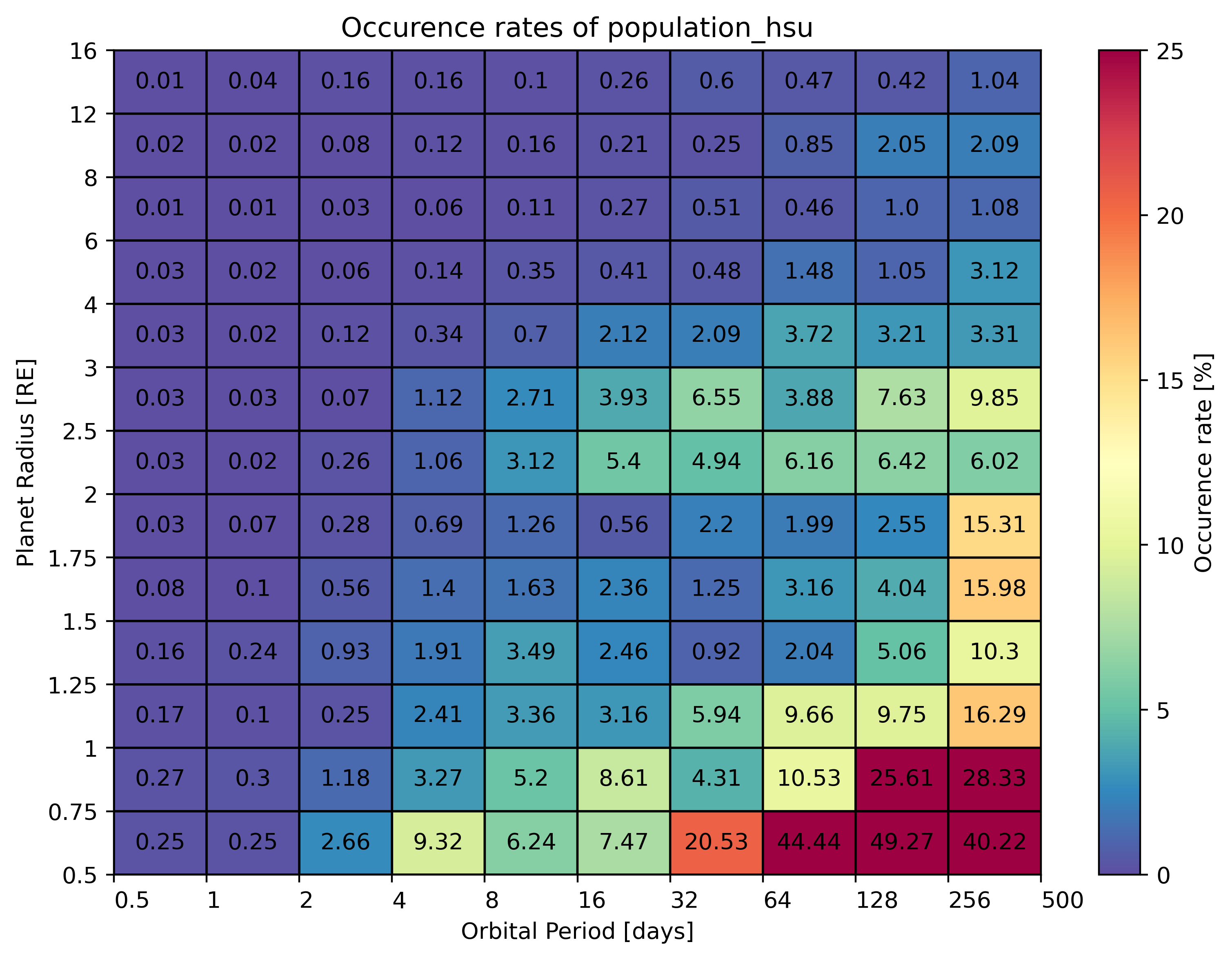}
        \includegraphics[width=0.45\textwidth]{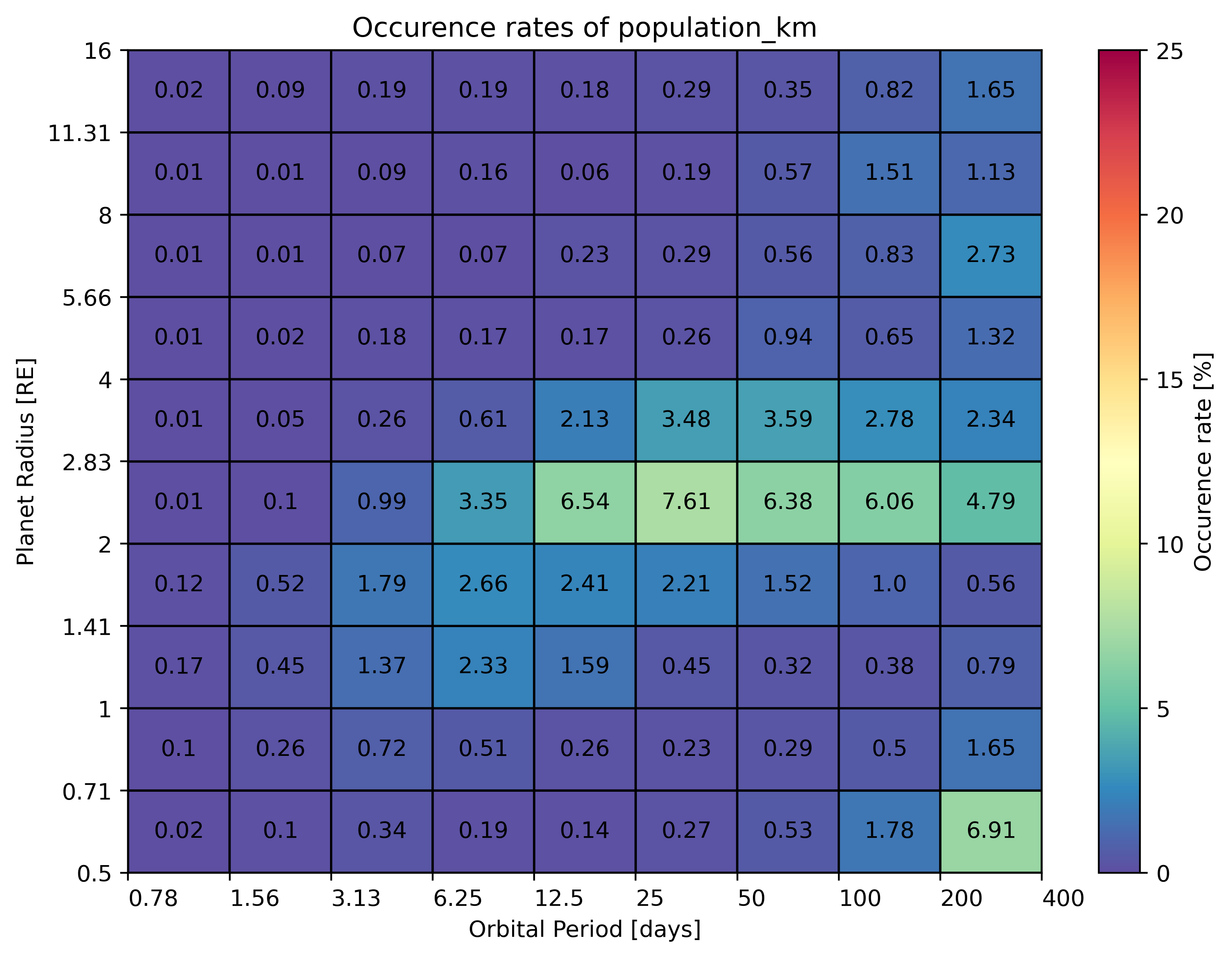}
    \caption{Inferred planet occurrence rates (color code) for the NGPPS76 5e9 population model (top) and mean values from 100 iterations for Hsu19 (middle) and KM20 (bottom panel), where we used the grid in orbital period and planet radius as in the original work, but the same color-coding for all three data sets.  }
        \label{fig:population_comp}
\end{figure}

In this section, we describe how we translated the planet occurrence rate into a planet population model similar to the NGPPS data set.

Formally, the occurrence rates of \citet{hsu2019} and \citet{kunimoto2020} are provided as mean values, and their $1\sigma$ uncertainty ranges over a set of cells. Each 2D cell is bounded by a lower and an upper value of $R_p$ and $P_{\rm orb}$. While the $R_{\rm p}$ and the $P_{\rm orb}$ grids can be unequally spaced, the 2D grid is rectangular. For each individual cell, the $1 \sigma$ uncertainties are the maximum values of a possibly asymmetric occurrence rate distribution in that cell.
In order to not only recover the mean but also represent its uncertainty in the planet population, we approximated the occurrence rate distributions of each cell as a symmetric Gaussian with $3\sigma$ upper and lower limits. 
For each cell, occurrence rate values were picked randomly until 100 values were admitted in a way to preserve a Gaussian distribution. With 100 draws, we represent the original mean values to within a few percent and the original $1\sigma$ uncertainty to within 10\%,

Given 1\,000 stars, an occurrence rate of 1\% means that there are 10 planets around these stars on average within the bounding $R_{\rm p}$ and $P_{\rm orb}$ values of that cell. Our aim is to simulate a relatively large number of planets to ensure that that the cells are approximately equally populated. Therefore, we used three different numbers of input stars per cell. For occurrence rates below 0.2\%, we took 50\,000 stars (yielding a maximum of 100 planets per cell), for rates below 2\%, we took 10\,000 stars (yielding between 20--200 planets per cell, and above 2\%, this yielded 1\,000 stars ($N_p>20$). For cells with only an upper limit, we assumed a uniform occurrence rate distribution between 0 and the upper-limit value, and we represented the population by 2\,000 stars. 

An occurrence rate $o$ in percent and a number of stars $N_s$ yields the number of planets populating that cell as $N_p=0.01\times o\times N_s$. These planets were assigned with a radius value assuming a uniform likelihood and a $P_{\rm orb}$ value assuming a uniform log likelihood within the bounding  $P_{\rm orb}$--$R_p$ ranges of that cell. Finally, each planet was randomly assigned with a starID between 1 and the $N_s$ value for that cell. This implies that we assumed random planet architectures.
For the subsequent calculation of the PPY, we kept track of the cell scaling factor $N_s$.
As a result, we obtained a table of planets with the starID, $R_p$ , and $P_{\rm orb}$ value. 

This process was repeated 100 times for the 100 occurrence rate representations of each cell, yielding 100 tables. Together, they built the population model given the input occurrence rate diagram. The 100 tables can also be viewed as 100 snapshots of sky observations in time. From the 100 tables, we can calculate the mean occurrence rate of each cell. These  values are displayed in Figure \ref{fig:population_comp} for HSU19 and KM20, respectively.

\section{Input stellar samples}\label{sec:stars}

The PLATO mission has been designed to monitor a large sample of $\gtrsim$ 245\,000 K7-F5 dwarf or subgiant stars of magnitude 13 or brighter, observed at 500-1000 nm wavelengths. Out of this P5 sample, a subset of $\gtrsim$ 15\,000 stars shall be observable at a photometric noise of 50 parts-per-million in 1h (ppm/h) or better; the subset with magnitude 11 or brighter forms the P1 sample (red book). 
The figure of 245\,000 targets assumes two long-duration observations, possibly separated into the northern and southern hemispheres. According to the nominal mission design, each of the fields will be monitored for transit events over a duration of two years. This is the 2$+$2 mission scenario (red book). An alternative scenario could be to stare at one field for three years and at several other fields for a shorter period of time (e.g., two months) in order to cover a larger portion of the sky during the four-year prime mission (the 3$+$1 mission scenario; red book).

A compilation of 2\,378\,177 stars fulfilling the above criteria and distributed throughout the sky is provided in the all-sky PLATO input catalog asPIC \citep{Montalto21}.
However, only the fraction of stars that falls within the fields of view will actually be observed by PLATO. The PLATO FsoV are of squarish shape and cover about 2\,000 deg$^2$ of the sky  \citep{Nascimbeni22,pertenais2021}. Approximating this shape by a square would yield a solid angle of 49 deg$^2$ according to \cite{Nascimbeni22}, which is 5.15\% of the sky.  Assuming a uniform distribution of the stars on the sky, we can estimate that each FoV contains 122\,476 asPIC stars. This is consistent with the minimum requirement of 122\,500 stars at mission level, although the eventually chosen fields in the 2+2 mission scenario may contain 20\% more stars \citep{Nascimbeni22}.

We  sorted the stars into magnitude V bins of 5-8, 9, 10, 11, 12, and 13 by rounding the values for \textit{gaiaV} provided in the catalog without accounting for their uncertainties, which are around 0.2 \citep{Montalto21}. 

We determined  the spectral class of each star from their $T_{\rm eff}$ values provided in the asPIC, neglecting the provided $1\sigma$ uncertainty $\sigma_{\rm Teff}$ (case $0\times\sigma_{\rm Teff}$). We used the relation between $T_{\rm eff}$ and spectral class for main-sequence stars of \citet{PecautMamajek13} as presented in their Table 5. If we are to collect KGF stars, we need to define boundaries between K0/G9 and G0/F9 stars, respectively.

Linear interpolation between the classes K0 and G9, which are separated in $T_{\rm eff}$ by 30 K \citep{PecautMamajek13}, yields a G to K transition at 5\,310 K. Linear interpolation between the classes G0 and F9, which are separated in $T_{\rm eff}$ by 120 K, \citet{PecautMamajek13}, yields an F to G transition at 5\,980 K.  To sort the stars into spectral class, we used these sharp F to G and G to K transitions. The respective upper limit for F-type stars of 7\,320 K and the lower limit of K-type stars of 3865 K is not important here as the asPIC only contains stars between K7 and F5. 

Neglecting the uncertainty $\sigma_{T\rm eff}$ induces an uncertainty in the star count. The individual 1$\sigma$ uncertainties of a star are around 227 K \citep{Montalto21}. Taking them into account (case $1\times\sigma_{\rm Teff}$) and averaging over 100 iterations, we obtain  a 1$\sigma$ uncertainty in the number of stars in each class of about 0.1\% for magnitudes 13 to 12, which increases to 1\% for the less abundant magnitude 8 stars. 
However, the mean values for the star numbers in each class differ from those obtained above, where we neglected $\sigma_{\rm Teff}$, by more than $3\sigma$ in some cases; see Table \ref{tab:stars_in_pic}. These differences are more pronounced the more unequal the number of stars in adjacent classes are, and they arise because a higher number of stars are distributed down the slope than upward.

To address a maximum uncertainty in the F to G and G to K transitions, we shifted the first by $+120$ K and the second subsequently by $-30$ K. The first modification would decrease the number of F stars over all magnitudes by up to 23\% and increase the number of G stars by 22\%. The second modification would increase the number of G stars by 3\% and decrease the number of K stars by 7\%. While spectral classes do not enter the computations of the PPY, they are important when the PPY for a certain stellar class is to be determined. 

For comparison, \citet{kunimoto2020} adopted values for F to G and G to K transitions of 6\,000 K and 5\,300 K, respectively. \citet{Montalto21} used a $T_{\rm eff}$ value of 6\,510 K (3\,870 K) that is representative of F5 (K7) stars \citep{PecautMamajek13} to limit their sample to K7--F5 stars.

In Table \ref{tab:stars_in_pic} we list the input stars obtained in this way before taking their 5.15\% fractions. To sort the spectral type by the effective temperature, we interpolated linearly in the F/G and G/K transition values of \cite{PecautMamajek13} and neglected the uncertainty in $T_{\rm eff}$ ($0\times\sigma$). For comparison, we show the included case  ($1\times\sigma$).

\begin{table}[h!]
        \centering
        \caption{Number of stars in the asPIC catalog according to our sorting}
        \begin{tabular}{ ccrrr  }
                \hline
                $n\times\sigma_{\rm Teff}$ & magV & F & G & K \\
                \hline
                0 & 8  &   5\,703 &   3\,231 &      909 \\
                0 & 9  &  18\,492 &  11\,749 &   3\,412 \\
                0 & 10 &  56\,423 &  41\,524 &  12\,431 \\
                0 & 11 & 155\,705 & 133\,336 &  43\,590 \\
                0 & 12 & 386\,006 & 407\,392 & 142\,710 \\
                0 & 13 & 352\,526 & 438\,531 & 160\,955 \\
                1 & 8  &   5\,471 &   3\,396 &      986 \\
                1 & 9  &  17\,820 &  12\,168 &   3\,665 \\
                1 & 10 &  54\,509 &  42\,423 &  13\,447 \\
                1 & 11 & 150\,642 & 134\,561 &  47\,428 \\ 
                1 & 12 & 379\,172 & 398\,879 & 158\,057 \\
                1 & 13 & 352\,329 & 419\,237 & 180\,447 \\
                \hline
        \end{tabular}\\
        \label{tab:stars_in_pic}
        \flushleft
\end{table}

Taking a fraction of 5.15\% of the stars of the asPIC as done in this work is equivalent to assuming a random pointing direction of PLATO. In Table \ref{tab:stars_in_lop1} in the appendix we compare this stellar sample from a random pointing direction with the pointing directions of LOPN1 and LOPS1 defined in \cite{Nascimbeni22}.

While the LOP1 fields contain 26--30\% more stars than the P5 requirement, we find 32--35\% more stars in these pointing directions that satisfy the P5 sample criteria. We also find that these fields contain disproportionally more F stars and more faint stars than the average sky field.
As at the current stage the final pointing direction is not yet fixed, we consider a random pointing a more robust approach than using a specific field for estimating the PPY.

\begin{figure*}[htb]
\centering
\resizebox{\hsize}{!}
{\includegraphics[width=0.98\textwidth]{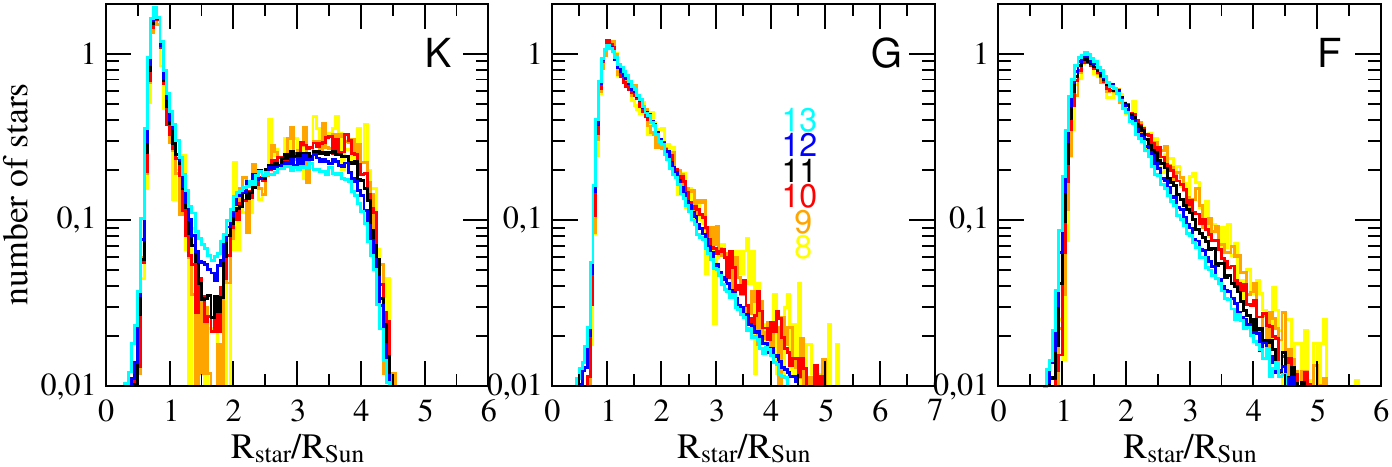}}
\caption{Histograms of the asPIC stars normalized to 1 and sorted for spectral class and apparent magnitudes between $m=8$ to 13 (color code as labeled). }
\label{fig:histoPIC_all}
\end{figure*}

Finally, we binned the stars into about 100 radius bins of constant width 0.05 $R_{Sun}$. Figure \ref{fig:histoPIC_all} shows the normalized  histograms of stellar radius after our sorting for stellar class and magnitude. 

\section{Detection efficiencies}\label{sec:DE}

The planet yield of a transit survey such as \textit{Kepler} or PLATO depends on several factors, including the detection efficiency of the search methods applied in the data pipeline~\citep[see, e.g.,][]{christiansen2020measuring}.

The detection efficiency can be parameterized as a function of the S/N. The signal is the measured transit. It is characterized by the transit depth (proportional to the radius ratio), its duration (proportional to the orbital period), and the number of observed transits (constrained by the observational strategy). The noise includes photon noise from the target, but also residuals from instrumental noise and stellar variability.

The  PLATO space telescope is equipped with 24 normal cameras. They are separated into four groups, which overlap each other in their viewing angle. When detection efficiencies are calculated, it is important to estimate how many cameras observe any given target.

The PLATO payload is required to achieve an overall reliability >79\% at the end of the extended mission lifetime (6.5 years). In order to account for the aging of the instrument and possible failures of hardware components, we defined two scenarios. At the begin-of-life, the instrument has full performance and 24 operational cameras. At the end-of-life, the photometric performance has degraded (e.g., because of lens darkening or because of radiation damage of the CCDs) and up to 2 cameras might have failed and may not deliver any data. Consequently, the free parameters added to our detection efficiency model are i) eight different numbers of cameras for a begin (24, 18, 12, 6; BOL) and end-of-life (22, 17, 11, 5; EOL) scenarios, ii) a stellar magnitude range between V 8 and 13, iii) the mission durations (0.1667, 1, 2, 3, 4, 6, and 8 years), and iv) the stellar types (F, G, K, and M).

\subsection{Relation between detection efficiency and signal-to-noise ratio}

The detection efficiency can be approximated as a simple function of the S/N with a finite-size threshold. Below the threshold, no planets are detected. Above the threshold, a certain fraction of planets is detected. In between, a smooth transition can be created by linear interpolation (see ~\citealt{jenkins1996matched,jenkins2010overview}, but also \citealt{hsu2019,kunimoto2020,christiansen2020measuring}).

For {\it Kepler}, \cite{fressin2013} suggested that the threshold should be between S/N = 6 and 16, where above the latter, 100\% of the planets should be detectable.  These limits are more conservative than previous calculations from the PLATO definition study report, which used an upper boundary of S/N = 10. For consistency with the previous PLATO estimates, we adopted the values from the red book. 

\subsection{Signal and noise models}

The S/N is the ratio of the transit depth $\delta,$ functioning as the received signal, and the noise $N$. The transit depth is calculated by
\begin{equation} \label{eq:delta}
        \ \delta = \frac{R_p^{2}}{R_{\ast}^{2}}
,\end{equation}
where $R_{*}$ is the stellar radius. Other geometrical parameters such as impact parameter or limb darkening, which influence the transit depth, are not considered in this work.

The noise can be addressed with the help of the effective combined differential photometric precision CDPP$_{\rm eff}$ , as outlined below. The noise 
\begin{equation} \label{eq:noise}
        N = \frac{CDPP_{\rm eff}}{\sqrt{t_{\rm transit}}\sqrt{N_{\rm transits}}}
\end{equation}
is given in ppm/h, but is dimensionless.
The number of transits $N_{\rm transits}$ is calculated with the mission duration $t_{\rm mission}$, the orbital period of the planet, and the transit duration $t_{\rm transit}$,\\
\begin{equation} \label{eq:6}
\  t_{\rm transit} = \frac{P_{\rm orb}R_{\ast}}{a \pi}
\end{equation}
\begin{equation} \label{eq:a}
\  a = \sqrt[3]{\frac{P_{\rm orb}^{2}G M_{\ast}}{4 {\pi}^2}}
.\end{equation}
The semi-major axis $a$ is calculated with Kepler's third law (Eq.~\ref{eq:a}), where $G$ is the gravitational constant, and $M_{\rm *}$ is the mass of the star. The mass of the planet $M_{p}$ is neglected here because $M_{p} \ll M_{\rm *}$. 
For large numbers of transits, a common approximation is
\begin{equation} \label{eq:approx}
\  N_{\rm transits} = \frac{t_{\rm mission}}{P_{orb}}
.\end{equation}
We requested a minimum number of transits of two before a signal was classified as a transiting planet. For instance, assuming a planet has $P_{orb} = 1.5 yr$ and $t_{\rm mission} = 2 yr$,  equation
\ref{eq:approx} yields $N_{\rm transits} = 1.3\bar{3} < 2,$ and thus the planet would be classified as a nondetection, although two transits could take place within two years. Therefore, we also took the timing of the first transit, $t_{0}$ , and the transit duration into account. As $t_{0}$ is not known, we generated 100 random numbers with a uniform distribution for $t_{0}$ within the interval 0--$P_{\rm orb}$. 
If 
\begin{equation} \label{eq:5}
\  \frac{t_{\rm mission}}{t_0+P_{orb}+t_{\rm transit}} > 1,
\end{equation}
the number of transits increases to $N_{\rm transits}+1$.
This occurred in $n_{+1}$ cases. We assigned the probability of $n_{+1}$/100 of observing one additional transit. We verified that the resulting probability is well converged for the choice of 100 random numbers.

The consideration of the timing of the first transit is particularly relevant for long-period planets such as Earth around Sun-like stars, which  have only a small number of transits within a few years of uninterrupted observation time.

\subsection{CDPPeff and the noise model}

The PLATO mission consortium has developed tools and models to assess the expected performance of the PLATO instrument. They can be used to predict the measured flux and the expected noise at pixel level and at mission level for a given star.
The noise budget of the PLATO light curves can be approximated with a model that includes  photon noise from the target, spacecraft jitter, and read-out and background noise. 
An additional noise parameter is the stellar variability $\sigma_{\ast}$.
This simplified model is appropriate for estimating the detection efficiency for a planet detection.
We express the noise budget as the CDPPeff in ppm/h (the noise in ppm for a signal that is averaged over 1h), although its dimension is $1/\sqrt{\rm time}$. The relevant timescale is represented by the transit duration in hours, \\
\begin{equation} \label{eq:cdpp_eff}
CDPP_{\rm eff} = \sqrt{\sigma_{\rm jitter}^2 +  \sigma_{ph}^2 + \sigma_{\rm readout}^2 + \sigma_{\ast}^2}\:.
\end{equation} 
The first three terms are collectively labeled CDPP, so that
\begin{equation} \label{eq:cdpp_eff2}
    CDPP_{\rm eff}^2 =  {\rm CDPP}^2 + \sigma_{\ast}^2\:.
\end{equation}

Jitter describes the noise that is produced by small-amplitude high-frequency movements of a satellite \citep{drummond2006jitter}. This error source affects the geometric accuracy of high-resolution satellite imagery \citep{pan2017satellite}. For the Transiting Exoplanet Survey Satellite (TESS), \cite{ricker2014transiting} pointed out that the most important source of systematic noise is due to jitter, while other noise sources decrease with the brightness of the observed star. For PLATO, jitter is estimated to be the dominant noise source at V < 9 (Cabrera et al. in prep.). We write the jitter noise as $\sigma_{\rm jitter}=k_{\rm jitter}$ with $k_{\rm jitter}=10$ ppm/hr (Cabrera et al. in prep.).

The CCDs work by amplifying the voltage induced by the electrons that were lifted into the conduction band after the absorption of photons in an array of pixels. The accumulated charge of the electrons in a pixel is read out before the pixel can be exposed to the light source again. However, the measurement of the electron charge and thus the number of electrons is imperfect. This instrumental noise source is called readout noise and is given as the number of electrons \citep{richmond2004readout}. We considered readout noise to be mixed with background noise (caused by background stellar contaminants or stray light) up to a value of $k_{\rm readout}=4 \times 57.8$ electrons (Cabrera et al. in prep.). The values for $k_{\rm jitter}$ and $k_{\rm readout}$ may be subject to change in the course of progress in the understanding of the PLATO instrument performance. 
The value of $k_{\rm readout}$ is provided in electrons, but it has to be combined with the other quantities that at system level are presented in ppm. For the estimate of the impact of the readout noise at system level, we assumed that it is not correlated in time or among cameras, hence
\begin{equation} \label{eq:sigma_readout}
\sigma_{\rm readout}^2=\frac{k_{\rm readout}^2}{f(m)^2}\times\left(N_{\rm cams}\:\frac{3600s}{25s}\right)
.\end{equation} 
Electronics readout and background noise are the leading noise source at the faint end (V>13).

Photon noise is a fundamental form of uncertainty in the measurement of light, inherent in the quantized nature of light \citep{hasinoff2014photon}. As emission of photons from a star is a spontaneous process, the uncertainty of the number of photons hitting the detector increases with the square root of the signal, which is the flux $f$ of incoming photons per unit area. The photon noise is expressed as a noise-to-signal ratio $N/S = \sqrt(f)/f = 1/\sqrt{f}$ so that $\sigma^2_{ph} = 1/f$. This stellar incoming flux changes with apparent magnitude $m$ as 
\begin{equation} \label{eq:flux_model}
\ f(m) = f_{mag11}* 10^{-0.4(m-11)}
,\end{equation}
where $f_{mag11}$ is our reference point for an magnitude 11 star. The electronic devices of the spacecraft are designed to yield an accuracy of CDPP = 50 ppm/h for an $m=11$ star observed by 22 cameras (EOL) simultaneously. We ensured this requirement by the choice of $f_{mag11} = 1.6\times 10^5$ (Cabrera et al. in prep.).

The stellar variability $\sigma_{\ast}$ was set to 10 ppm in 1h following the approach for Kepler, with its white noise being 10 ppm for a 12th magnitude G2V planetary target star \citep{van2016kepler}.

\begin{figure}[h]
        \includegraphics[width=0.5\textwidth]{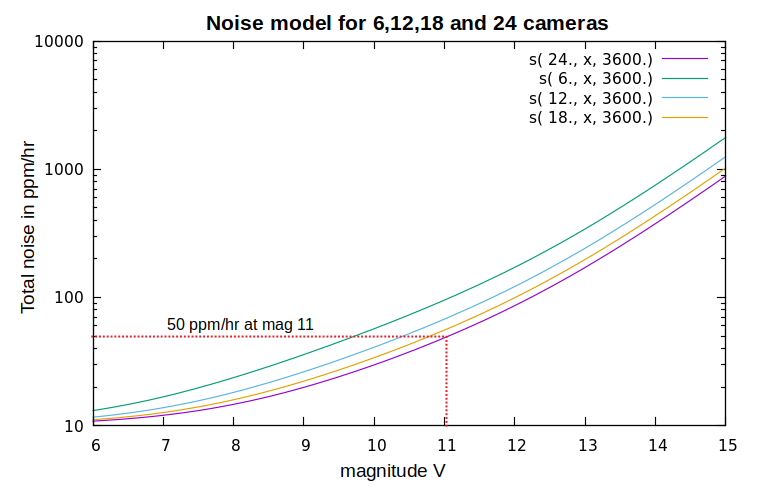}
        \caption{Noise model for a begin-of-life scenario with four camera configurations (6, 12, 18, and 24 cameras). The dotted red line shows the reference point of 50 ppm/h for a magnitude 11 star.}
        \label{fig:noisemodel_magV}
\end{figure}

Figure \ref{fig:noisemodel_magV} illustrates the noise CDPP in dependence on the apparent stellar magnitude and the assumed number of cameras observing the target.

\begin{figure}[h]
        \centering
        \includegraphics[width=0.5\textwidth]{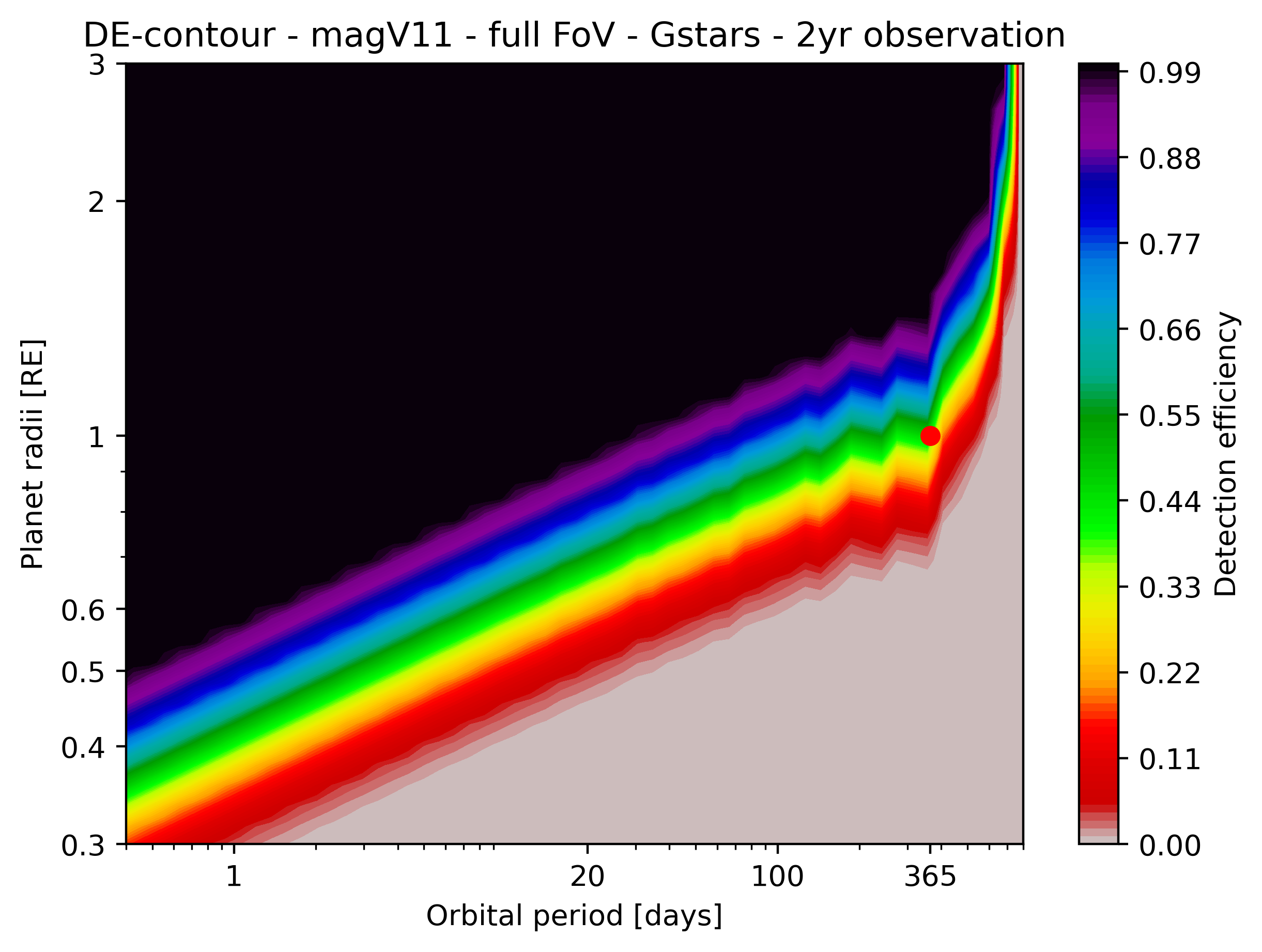}
        \caption{Example detection efficiency contour grid observing an 11 magnitude G star for two years with 24 cameras} 
        \label{fig:final_de_grid}
\end{figure}

Finally, we can calculated the $S/N$ and interpolated the detection efficiency (DE) from this. While in the simulation, we calculated DE values for the individual planet-star combinations, we show in Figure \ref{fig:final_de_grid} the detection efficiency on a grid of $R_p$ and $P_{orb}$ values for illustrative purpose  specifically for a magnitude 11 G star and an observation time of two years.

\section{Planet Yield for PLATO Estimator }\label{sec:PYPE}

The tool called Planet yield for PLATO Estimator is a Python program for the determination of the number of planets that can be detected, given a planet population model, a sample of stars that shall be observed, and the characteristics of the detection efficiency. The development of PYPE has been heavily inspired by and benefited from the open-source program Exoplanet Population Observation Simulator (EPOS) of \cite{mulders2018exoplanet,mulders2019exoplanet}, which we used initially. \\
The basic approach of PYPE is structured as follows: \\
(i). Defining a planetary population model. Orbital period values are converted into orbital distance and vice versa, assuming a planet mass $M_p=0$ unless the $M_p$ values are provided with the population model, which is the case for NGPPS. 
\\
(ii). Defining a stellar sample to be observed. If the number of stars to be observed ($N_{*,\rm obs}$) exceeds the number of systems with planets $N_{\rm sys}$  in the planet population model  by a factor of two or more, identical copies of the planetary systems are made to approximate $N_{*,\rm obs}$. The following steps are thus made for a number of stars $m\times N_{sys}$, $m\geq 1$. This idea has been adapted from the EPOS code. 
Here, we used 5.15\% of the number of stars from Table \ref{tab:stars_in_pic} and scaled the resulting PPY down to 5.15\% to account for the effective FoV of PLATO, as discussed in Sect. 3.\\
(iii.) 
The next step is to separate the transiting from the nontransiting planets. This step is purely geometric and calculates the probability of a transit using \begin{equation} \label{eq:geo_prob}
\ f_{geo} = R_{*}/a ,
\end{equation}
where $R_{*}$ is one of the binned $R_{*}$ values. If the geometric transit probability $f_{geo}$ is greater than a random variable $\epsilon\,[0,1]$, the transit occurs.  Unlike the EPOS code, which does take care of mutual inclinations between planets in the same system, we neglected information on planetary architectures. This perhaps unnecessary simplification matters when only a few systems are observed, but it will average out when many systems are observed. We may therefore underestimate the uncertainty in the PPY when the PPY is low. 
\\
(iv). We separated detected and nondetected transiting planets using the detection efficiencies of PLATO (\S \ref{sec:DE}), which adopt values between $f_{det}=0$ and 1 (100\%). If  a random value  $\epsilon\, [0,1]$ is lower  than $f_{det}$, the planet is counted for the PPY. 
In steps (iii) and (iv), the stellar radius distribution matters most.
\\
(v) While we ensured that the planets have semi-major axes $a$ greater than $1\, R_{*}$, observed ultrashort-period planets are found to be limited to $a/R_{*}>2$ \citep{Adams21} and to be small. Larger gaseous planets may be more readily subject to Roche-lobe overflow, and disintegrate. By neglecting these constraints, we slightly overestimate the PPY for planets at small orbital distances.
\\
(vi). The obtained PPY is scaled from the simulated number of stars, $m\times N_{\rm sys}$, to $N_{\rm obs}$. The PPY obtained from PYPE in this way is valid for a given single-mission duration, number of cameras, stellar class, stellar magnitude, and a single input population model.\\
(vii). After-PYPE:
The complex lotus-shaped FoV is taken into account. 
PLATO does not observe the whole FoV with all cameras simultaneously. Instead, different fractions of the field are observed with 6, 12, 18, or 24 cameras.
We multiplied the PPY  with a factor that stands for the percentage of the FoV observed by each number of cameras. For example, only 13.79 \% of the FoV are observed with 24 cameras, meaning that the results for the computation with 24 cameras was multiplied by 0.1379.\\
(viii). The above steps were repetitively executed for different mission durations, spectral class, and stellar magnitude.\\ 
(ix) The entire process was repeated $10\times$ (NGPPS) or $100\times$ (HSU19, KM20) in order to estimate the $1\sigma$ uncertainties and the mean values of the PPY.\\
(x) Finally, the PPY was sorted into $R_p$ and and $P_{orb}$ bins. We adopted the classification of planets according to their radius similar to \cite{fressin2013}(s. Table 2), which precedes the finding of a radius gap at 1.5--2.0 in the $R_p$--incident flux relation \citep{Fulton17}.
\begin{table}[h]
\caption{Planet radius bins for the PPY}
\centering
\renewcommand{\arraystretch}{1.2}
\begin{tabular}{c|c}\hline
$R_p$-range ($\RE$) & Planet type \\\hline
6--15 & Jupiters\\
2--6 & Neptunes \\
1.25--2.0 & Super-Earths \\
0.8--1.25 & Earths \\
0.5--0.8 & Sub-Earths
\end{tabular}
\end{table}

\section{Results}\label{sec:results}

Our calculations with PYPE for an estimate of the PPY have led to an immense amount of data regarding the number of detected planets as a function of mission duration, stellar radius, stellar magnitude, and planet type. The following graphs in this section display only a small portion of the results. Figures similar to Figure 8 displaying the results for other parameter choices as well as comprehensive data tables of PPY as a function of magV, mission duration, stellar class, and planet type can be found in more detail in the \href{https://dx.doi.org/10.6084/m9.figshare.21394005}{Online repository}\footnote{https://dx.doi.org/10.6084/m9.figshare.21394005}.

\subsection{Planet yield dependence on free parameters}\label{sec:results1}

\begin{figure}[h]
        \includegraphics[width=0.48\textwidth]{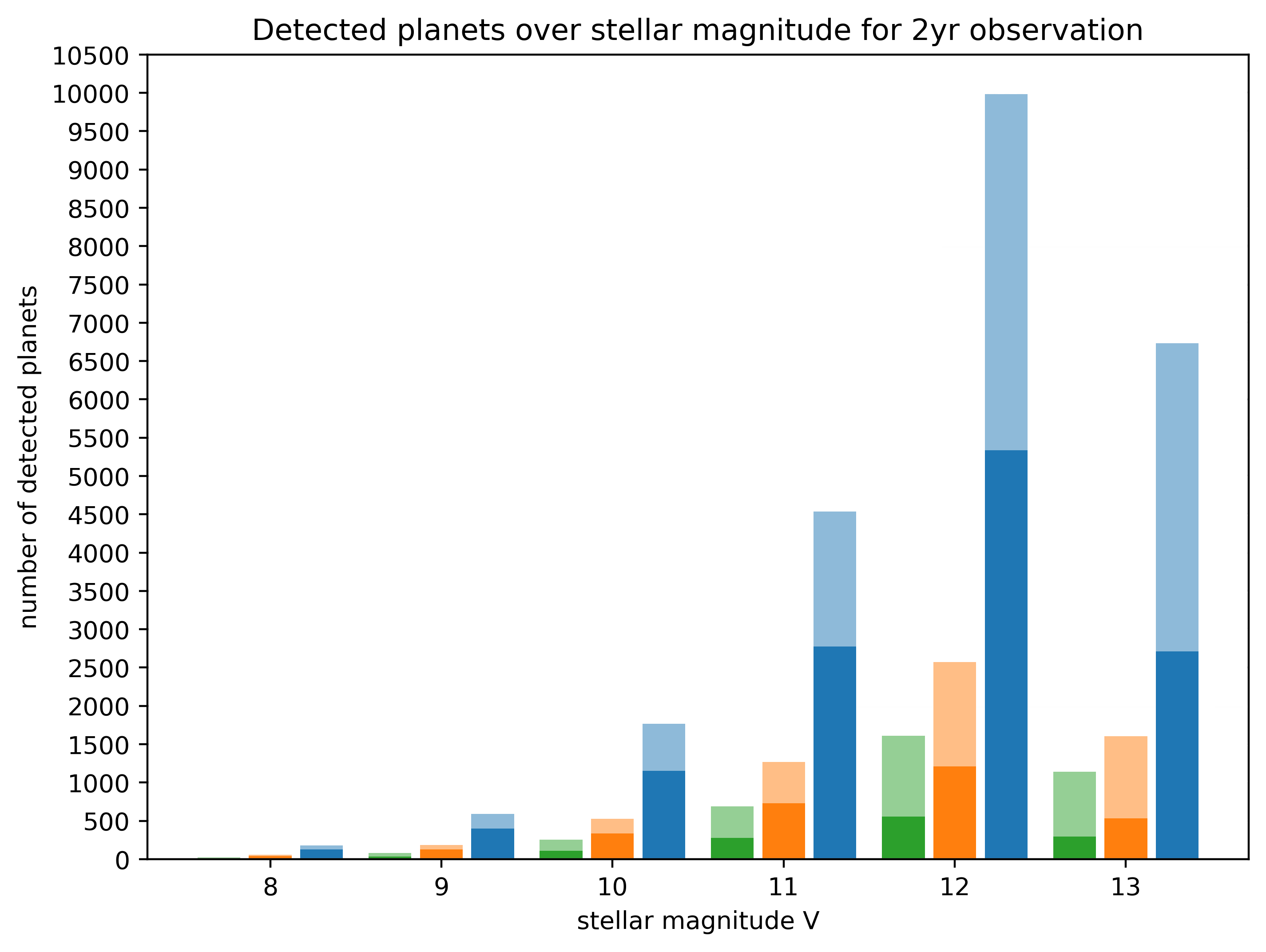}
        \caption{Number of detected planets for given magnitude range for NGPPS (blue), HSU19 (orange), and KM20(green). The darker bars show planets <2 $\RE$ , and the lighter bars show planets $>$2$\RE$. Every bar represents the PPY for BOL.}
        \label{fig:ndet_magv}
\end{figure}

In Fig.~\ref{fig:ndet_magv} we plot the PPY in dependence on the stellar magnitude for a 2+2 mission scenario and BoL. We find that most planets will be detected around magnitude 12 stars, followed by stars of magnitude 13 and 11. The relative behavior of PPY over magnitude V is similar for all three population models (green, orange, and blue); strong differences occur in absolute numbers. The majority of planets are detected around fainter stars (V $>$ 11), while the RV-feasible planets (V $\leq$ 11) make up between 38\% and 49\% of detections. The PPY ratios of these two magnitude groups are 2.4 (NGPPS), 2.1 (HSU19), and 2.6 (KM20).

\subsection{Detectable planet occurrence rates}\label{sec:results2}

\begin{figure*}[th]
    \centering
        \includegraphics[width=0.99\textwidth]{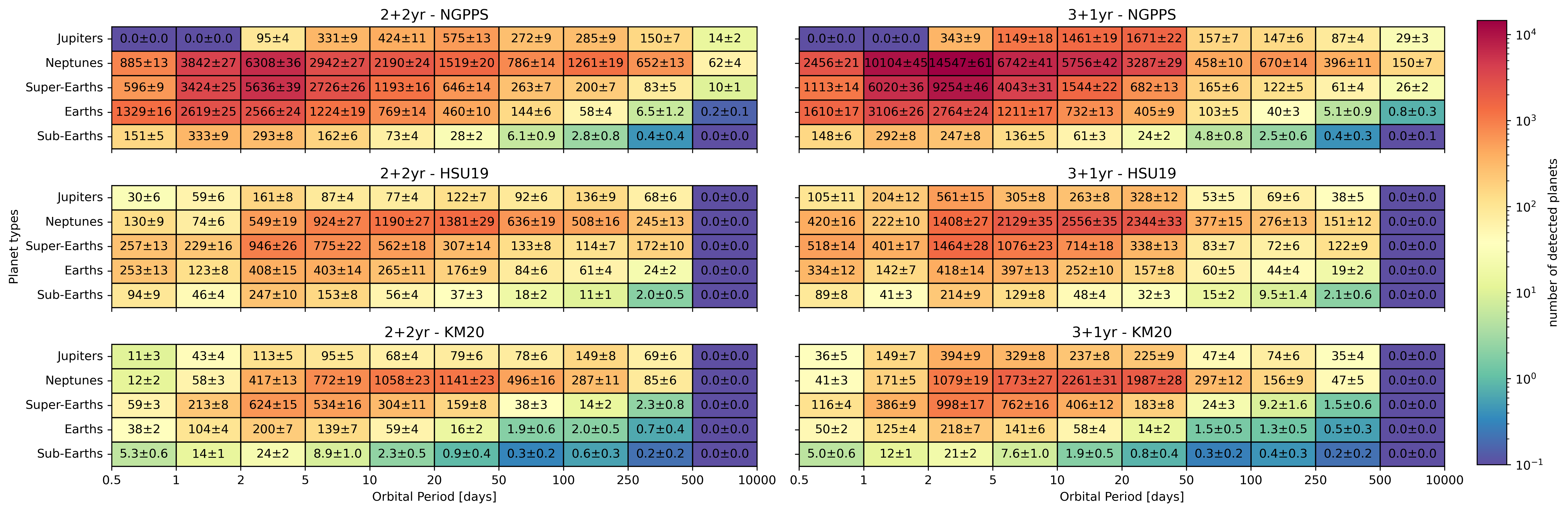}
        \caption{PPY as a function of binned planet size and binned orbital period for NGPPS (top row), HSU19 (middle), and KM20 (bottom), and for a 2+2 mission scenario (left) or a 3+1 mission scenario (right).}
        \label{fig:ppy_occ_22u31}
\end{figure*}

Figure \ref{fig:ppy_occ_22u31} shows the PPY subdivided into planet type and orbital period bins for the $2+2$yr and $3+1$yr mission scenarios. For most bins, the 3+1yr case yields a higher planet count than the 2+2yr case. The larger a planet and the shorter its orbital period, the more strongly the PPY benefits from the 3+1 case. This is because in the 3+1 case, seven different fields are observed and thus a larger number of different planets, which can be rather quickly detected with  two transits when they are on short orbital distance and large. 

The opposite holds for smaller and more distant planets with an orbital period greater than 50 days (i.e., approaching two months). Clearly, in the +1 short-duration case, Earth-size planets in the HZ cannot be found when the FoV is changed every two months. 
On the other hand, smaller planets on long orbital periods benefit from a third year of continuous observations of the same field as compared to only two years. 
Together, these opposing effects lead to a $\sim$ 50\% higher PPY in the 2+2 than in the 3+1 case when the orbital period exceeds one month. 

The PPY between the NGPPS population and the HSU19-KM20 populations shows large  $R_p$, $P_{\rm orb}$ specific differences. For instance, with NGPPS, we estimate the detection of $\sim 6320$ hot Neptunes in 2-5 days, whereas with HSU19 and KM20, we estimate only $\sim 550$ and $\sim 420$, respectively.
For earths in the 250-500 days bin, which partially overlaps with HZs of GK stars, we see a different picture. There, the highest PPY is found with the HSU19 population in the amount of $24\pm 4$, followed by NGPPS with $\sim 6.6 \sim 1.0$, and KM20 with only $0.7\pm 0.4$ planets. Planet yields for orbital periods greater than 500 days are at zero for HSU19 and KM20 because the given occurrence rates do not cover this area.

\begin{figure*}[th]
    \centering
        \includegraphics[width=0.99\textwidth]{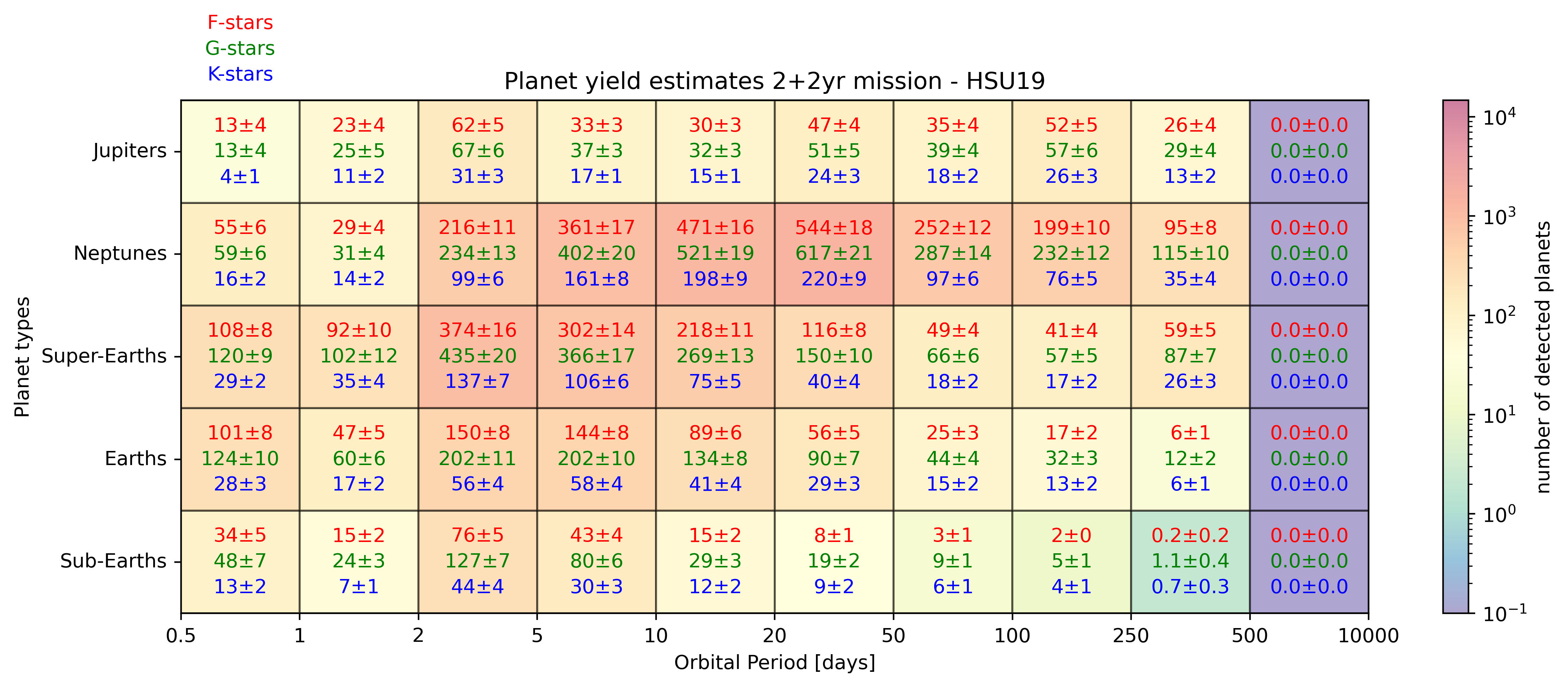}
        \caption{PPY as a function of binned planet size and binned orbital period for the 2+2yr mission scenario and the HSU19 population around F (red), G (green), and K stars (blue).}
        \label{fig:HSU19_2_2_FGK}
\end{figure*}

In Figure \ref{fig:HSU19_2_2_FGK} we show an example in which the PPY is subdivided according to stellar type. This particular example is for a 2+2yr mission scenario with the HSU19 population, but the same pattern is seen for every possible constellation of planet populations and mission scenarios. The tables for other parameter choices can be found in the \href{https://dx.doi.org/10.6084/m9.figshare.21394005}{Online repository}\footnotemark[3]. \\
Figure \ref{fig:HSU19_2_2_FGK} shows that G stars generally have the greatest number of planets flagged as detections, followed by F stars and K stars, except for small planets at wide orbits. F stars have the highest abundance in the PIC catalog, followed by G and K stars. However, the transit depth of small planets around F stars is lower than for K stars. The low signal for F stars can be compensated for by a low noise level, which can be achieved by multiple transits and thus is possible for short-period planets, but not for long-period planets. Thus for the latter, the deeper transit in front of K stars leads to more detections of small planets, even  though K stars are less abundant in the PIC catalog.  
Furthermore, for HSU19 and the 2+2 case, which we argued above is more favorable to detecting earths at wide orbits, we obtain a maximum of a dozen $\pm 2$ detected Earth-size planets around G stars in the 250-500d period bin.
While agreeing with the value of 11 detected Earth-size planets in the conservative HZ around Sun-size stars predicted by \citet{Heller22} for a 2+2 mission scenario, we add a word of caution as the assumptions behind both studies are very different. \citet{Heller22} assumed $2\times 7500$ bright bright (magV $\leq 11$) stars, corresponding to the P1 sample, and a higher true $\eta_{\rm E}$ of 37\%, whereas our estimate is derived from the larger but on average fainter P5 sample and a lower $\eta_{\rm E}$ of 16-28\%. Our estimated value decreases by 20\% when changing to a 3+1yr mission. 
For NGPPS, we find $3.7\pm 0.7$ , and for KM20 $0.3\pm 0.3$ for a 2+2yr case. \\
When we allow for a planet radius up to $2\RE$ (as for habitable planets in the PLATO definition study report), we estimate the planet yield around G stars to be $100\pm 7$, $42\pm 3,$ and $1.6 \pm 0.7$ for HSU19, NGPPS, and KM20 in the 250-500d period bin, respectively.

\subsection{Single-transit detections}\label{sec:results3}


The results presented so far were obtained under the premise that at least two transits
are observed by PLATO, in accordance with the definition study report.
However, single-transit events will occur. In this section, we quantify the increment    
in the PPY that arises from single-transit events. 

\begin{figure}[th]
\rotatebox{270}{\includegraphics[width=0.36\textwidth]{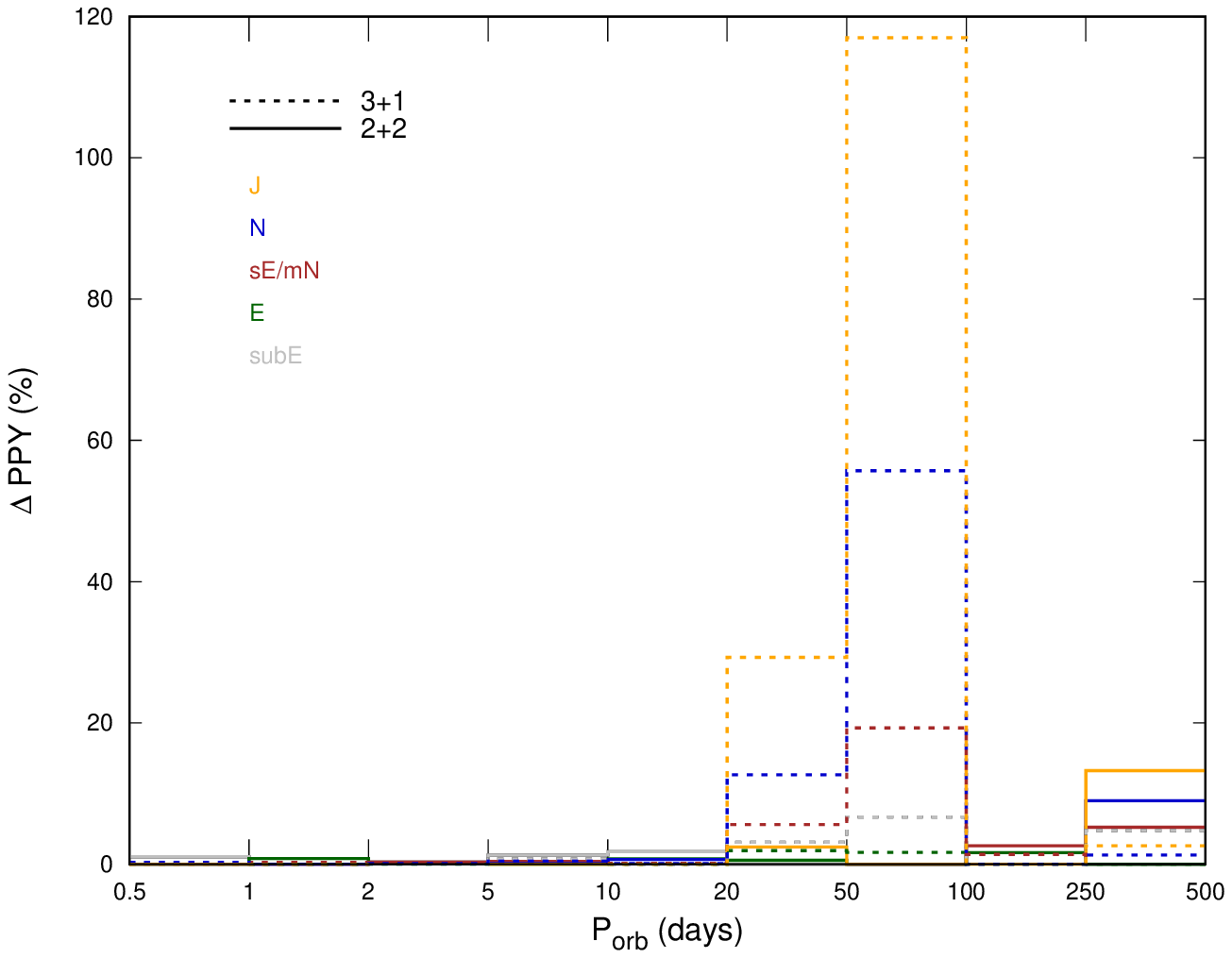}}
\caption{\label{fig:diffPPY}
Relative change in the statistical mean values of the PPY that results from the inclusion 
of planets that are observed to transit only once. Results are presented for the HSU19 planet 
occurrence rate and the 2+2 (solid) or the 3+1 (dashed) mission scenarios at BOL for Jupiters (J, yellow), Neptunes (N, blue), Super-Earths (sE/mN, brown), Earths (E, green), and Sub-Earths (subE, gray).}
\end{figure}

Figure \ref{fig:diffPPY} shows the relative increment of the statistical mean value 
of the PPY that results from the inclusion of planets that are observed in transit only once.

Small positive or negative fluctuations in the relative change can occur, which result 
from the $1\:\sigma$ uncertainties of the statistical means. These fluctuations are lower than 
1\% for most $P_{\rm orb}$--$R_p$ bins. Because negative values are solely due to this statistical uncertainty, 
only positive changes are shown.

A statistically significant change can occur if $0.5 < P_{\rm orb}/t_{\rm mission} <  1$ 
(case $i$) or if $P_{\rm orb} > t_{\rm mission}$ (case $ii$). In the first case, two transits might have been observed by PLATO, but due to unfortunate timing of the first transit, the second one
is missed. Half of the planets with such an orbital period will appear as
single-transit events on average, while the other half will appear with two transits, and will be detected or not detected according
to their DE. In the second case, PLATO can see one transit at most.

As expected, Figure \ref{fig:diffPPY} therefore displays an increment in the PPY when cases
$i$ or $ii$ apply. In the 2+2 mission scenario (solid lines), the increment adopts  
statistically relevant values only in the longest-period bin, where both cases apply. 
As larger planets have a better DE, their detectability with only one transit is higher, 
and thus their PPY benefits more than that of smaller planets. For Neptune- to Jupiter-size 
planets, the PPY increases by 9--13\%, given the input occurrence rate. It is clear that 
transit detections of extrasolar Neptune- and Jupiter-analogs in terms of $R_p$ and $P_{orb}$ 
can only occur via single-transit events.

The largest increments of up to a factor of two are obtained for the 3+1 mission scenario.
As they occur for $P_{\rm orb}$ within 20--100 days, these strongest increments entirely result 
from the step-and-stare phase. A much smaller increment of up to 5\% 
is seen in the 250- to 500-days bin, which results from the three-year observing phase.

This study emphasizes the importance of single-transit detections for the study of the 
architecture of Solar System analogs.
Further figures for different mission scenarios and planet populations can be found in the \href{https://dx.doi.org/10.6084/m9.figshare.21394005}{Online repository}\footnotemark[3].

\subsection{Planet yield of LOPN1 and LOPS1}\label{sec:results4}

\begin{figure*}[th]
        \includegraphics[width=0.99\textwidth]{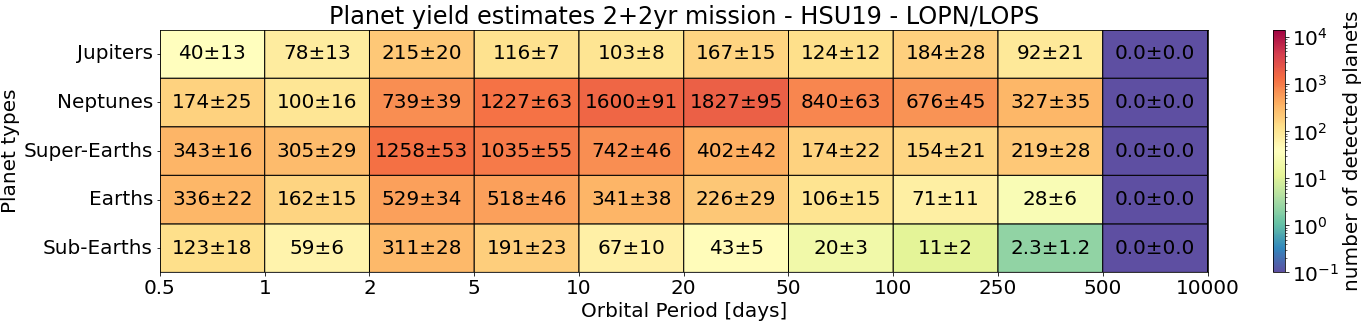}
        \caption{BOL-PPY as a function of binned planet size and binned orbital period for the HSU19 population and the 2+2yr mission scenario using the LOPN1 and LOPS1 stellar catalog.}
        \label{fig:HSU19_LOPN+LOPS}
\end{figure*}

While the definite FoV for PLATO has not yet been chosen, \cite{Nascimbeni22} has proposed two fields for the long-observation phases. One lies in the northern (LOPN1) and the other in the southern (LOPS1) ecliptic hemisphere. Taking the same pointings for PLATO as that study and cross-referencing it with the asPIC from \cite{Montalto21}, we obtain 165351 and 161631 stars for LOPN1 and LOPS1, respectively. These stars underwent the same treatment as the asPIC in section 3. The PPY was calculated for the HSU19 population for a 2+2 yr mission scenario as a 3+1yr mission would not qualify because the additional +1yr period would not be a long observation, but a step-and-stare phase that changes the pointings every two months. For the 2+2yr mission, the first 2yr calculation was made with LOPN1 and the second with LOPS1. The results in figure \ref{fig:HSU19_LOPN+LOPS} for the nominal pointings LOPN1/LOPS1 can be compared directly with the results from figure \ref{fig:ppy_occ_22u31} for a random pointing. The nominal pointings yield a larger number of planets, but it is not dramatically different, as the expected result of the optimization criteria used in the selection of the PLATO pointing fields (see \cite{Nascimbeni22} for details).

\section{Discussion}\label{sec:discussion}

The scenarios presented here primarily studied the current baseline observing duration of four years (as constrained by the ESA schedule planning) and the scenarios as addressed in the red book. 
We recall that PLATO is designed for a six-year observing duration and will have consumables for up to eight years of operations. 
To take advantage of these options, mission extensions can be applied for after launch of the mission. 
These scenarios deserve further studies that are currently ongoing.

\paragraph{NGPPS model.} 
With input occurrence rates a factor of a few higher than that of HSU19, NGPPS unsurprisingly predicts the largest PPY. Thus, the bulk PPY from NGGPS model appears to represent the high end of expectations for the PLATO mission. However, it necessarily makes a number of simplifying assumptions. Every planetary system in the NGPPS model is assumed to form and evolve around a Sun analog with a mass of 1 $M_{Sun}$. This does not reflect the variety of stars that will be observed. Planetary systems formed around stars of different mass and metallicity are observed to exhibit different architectures. Moreover, some planets may have formed by means other than core accretion, as indicated by the recent combined ALMA and HST observations of a few million years old young protostellar disk \citep{Currie22}. A pathway to understand planet formation could be to compare different predictions to the observed planet yield. 

\paragraph{HZ.} 
The location of an HZ is still a matter of ongoing debate as it depends on our understanding of the stellar flux \citep{Ahlers22}, the effect of clouds in the planetary atmosphere \citep{Leconte13}, and of geological processes such as plate tectonics and the temperature-stabilizing carbon cycle \citep{Kruijver21}, although the ability to sustain liquid water on the surface is commonly considered a necessary condition for habitability. \\
The HZ of a planet could be calculated with the effective temperature and luminosity of its host star and a model for the radiation transport in the planetary atmosphere \citep{Kopparapu_2013,Kopparapu_2014}. Uncertainties then result from the global mean properties such as atmospheric composition, in particular, the abundance of greenhouse gases, and from opacities. \citet{Kopparapu_2013} related the CO$_2$ abundance to incident flux at the top of the atmosphere. Conditions satisfying the criterion for liquid water at the bottom of the atmosphere result in a conservative habitable zone of 0.99-1.70 AU for the Sun. \citet{Kopparapu_2013} provided a relation to estimate the HZs of other stars. Calculating a specific habitable zone for every individual star \citep{Bryson21} would exceed the scope of this study.  Given the uncertainties, we are only interested in estimates here. Assuming representative values for the effective temperatures (5000, 5800, and 6500 K) and stellar radii (0.7, 1, and 1.3 $R_{\rm Sun}$), we can calculate the HZs of 153-358, 340-797, and 733-1678 days for KGF stars, respectively. With KM20 and HSU19, which only reach orbital periods of up to 400 and 500 days, respectively, the HZ around K stars is fully covered, while the G-star HZ is only partly covered and the F-star HZ is entirely missed. 

Using NGPPS, we estimate that PLATO can find one planet there when the monitoring duration is limited to three years. Given the closer-in HZ boundaries around K stars, we estimate that PLATO can detect a few (KM20) to a few dozen planets in the 0.8-$1.25\RE$ size range. These numbers increase quadratically if larger planets are admitted. \cite{Bryson21} considered planets up to $1.5\RE$ for their $\eta_{\oplus}$ estimate, while in the red book, radii of even up to $2\RE$ are considered.

\paragraph{Comparison to the red-book results.}
An earlier study (red book) predicted a total planet yield for planets around stars with magnitudes of V$\leq$13 to be $\sim$4\,600 and $\sim$11\,000 for the mission scenarios 2+2yr and 3+1yr. Our most conservative population KM20 predicts PPYs being respectively $\sim$70\% and $\sim$32\% higher than this. The NGPPS population exceeds the red-book estimates by a factor of 10 and 7.5. In the previous study, the estimates were based on the occurrence rates by \cite{fressin2013}, which are given there for similar planet types, but only an orbital period range of 0.8 to 85 days. This means that various assumptions regarding the occurrence rates of more distant planets have been made. Our estimates fill these unknowns with observational and theoretical values, showing that the previous assumptions were highly conservative.

\paragraph{RV feasibility.} The brightness of a star determines the possibility of constraining the bulk properties of exoplanets. The fainter the star, the lower the likelihood of providing accurate planet parameters for a detected exoplanet through RV measurements. Figure \ref{fig:ndet_magv} shows that the majority of detected planets are found for magnitude V>11 stars and are therefore too faint for their follow-up RV measurements. Nonetheless, the number of detections around V $\leq$ 11 stars is still impressive compared to the 1228 already found and confirmed exoplanets at V $\leq$ 11 (NASA Exoplanet Archive, May 23, 2022). An objective of PLATO is to detect planets around stars with brightness V $\leq$ 11. With the presented estimates, this goal will be achieved for one-third of the planets, and following RV measurements will be enabled.

\begin{figure}[h]
    \centering
        \includegraphics[width=0.48\textwidth]{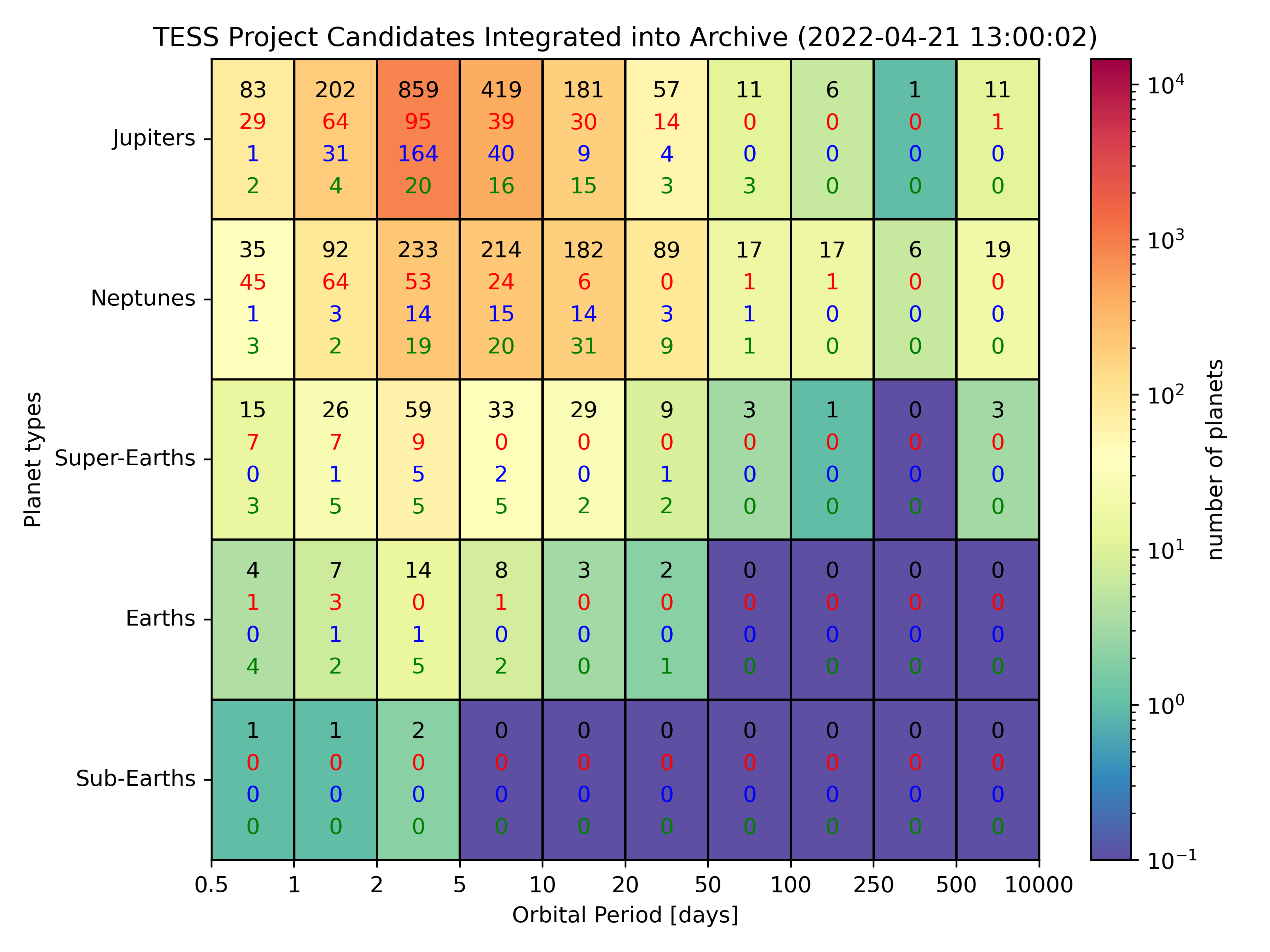}
        \caption{TESS exoplanets as a function of planet type and orbital period, separated into planet candidates (black), false positives (red), previously known planets discovered by other missions than TESS (blue), and confirmed planets verified through additional observations using other telescopes (green).}
\label{fig:tess}
\end{figure}

\paragraph{Rediscoveries.}
PLATO will detect several thousand planets. The question then is how many of these are already classified as candidates or confirmed planets by other space missions and ground-based observations. The 
\textit{Kepler} FoV could be within one of PLATOs long-duration FsoV, depending on the final pointing strategy chosen by the PLATO team~\citep{Nascimbeni22}, meaning that a decent number of \textit{Kepler} and K2 exoplanets will probably be rediscovered. However, TESS observed approximately 30\,000 $deg^2$ of the sky for a minimum of 27.4 days \citep{howard2015transiting}. Figure \ref{fig:tess} shows the TESS project candidates\footnote {https://exoplanetarchive.ipac.caltech.edu/} (by 2022 May 13) that are similar in planet type and orbital period bins to Figure \ref{fig:ppy_occ_22u31} and distinguished for the status of confirmation, which is either candidate, false positive, confirmed (meaning verified through additional observations using other telescopes), or previously known planets that were discovered by other missions before TESS.
As most of the TESS FoV is observed only up to 27 days, the majority of the candidates are found at short orbital periods; the TESS planet yield peaks for hot Jupiters. From the number of objects already evaluated to count as confirmed or a false positive, we estimate that 150 out of the 860 current hot-Jupiter candidate TESS planets may later be confirmed as planets. Together with the previously known hot Jupiters, the number of 314 previously known hot Jupiters is with a factor 0.5-3 about the same as our PPY estimate.  This means that a significant fraction of the hot Jupiters to be found by PLATO will probably be rediscoveries. It should be noted that planets discovered by TESS include planets around M stars, which are not included in the star sample studied in this paper (although PLATO will observe M stars as part of the P4 population; see \citet{Montalto21}). For an estimate of the TESS-expected yield through the prime and extended missions, see \citet{kunimoto2022}.

\paragraph{Single-transit events.}
Single-transit events can occur if $P_{\rm orb}>0.5 t_{\rm mission}$.  In the 2+2yr case, this applies to planets with $P_{\rm orb} > 1$ yr. In the respective bin of 250-500 days, an increment up to 14\% is indeed seen. 
Although this is small, it suggests that an analysis of  single-transit events may reveal planets at longer orbital distances that otherwise would be discarded, where the true occurrence rate is still highly uncertain. 

For the 3+1yr case, a strong PPY increase occurs at orbital periods of 20-100 days as it benefits from the step-and-stare phase (6 times two months) of this mission scenario. Lowering the $N_{\rm transits}$ threshold enables us to detect some planets that were previously discarded. The detection of Earth-Sun analogs does not seem to profit from this change as other parameters such as the planet radius seem to play a stronger role.

On the other hand, even a few detections at wide orbital distances would be valuable for building up a data base for the occurrence rate of planets at wide orbital distances. This will have to be set up over decades and by combining single-transit observations from various missions such as Kepler, TESS, or NGTS. Even   partial single-transit events from different observing programs might be combined that otherwise remained undetected. This may help to detect Earth-Sun analogs.

\paragraph{The LOPN1 and LOPS1 stellar fileds.} 
By comparing the PPY using the average stellar sample with that obtained with LOPN1 and LOPS1, we see that the former results amount to about two-thirds of the LOPN1/LOPS1 results. The number of stars that were used show the reason for this. Scaling the 2,374,623 stars in the asPIC down to 5.15\% (to account for the FoV size of PLATO in a given random pointing direction) and assuming a uniform distribution leaves us with 122,293 stars in each field. Observing two fields means 244,586 stars, which is about two-thirds of the 326,982 stars of LOPN1+LOPS1. This is an expected consequence of the field optimization criteria used in \cite{Nascimbeni22}.

\section{Conclusion}\label{sec:summary}

We have applied three different input occurrence rate estimates, 
information on the stellar distribution to be observed by PLATO from the
asPIC catalog, and a noise model to estimate the PLATO planet yield
as a function of planet size, orbital period, mission duration,
and stellar magnitude. Our conclusions from this study are listed below.\\
\\
\begin{enumerate}
\item
The largest unknown in our model is the true number of planets regarding their sizes and orbital periods. We implemented the combined knowledge from real observational data and the most sophisticated theoretical model up to date to solve this problem. Both \cite{kunimoto2020} and \cite{hsu2019} relied on observational data acquired by the NASA \textit{Kepler} mission. However, they covered only occurrence rates up to orbital periods of 400 and 500 days, respectively, while the core-accretion planet formation model NGPPS \citep{emsenhuber2020a} contains planets out to 2.7 yr. The bulk planet yields reach $\sim 4000$ (KM20), $\sim 6000$ (HSU19), and $\sim 24,000$ after two years of uninterrupted monitoring. These rather large differences suggest that the uncertainty in the current PPY estimates might be as large as a factor of a few. The uncertainty from the number of stars contributes less than 20\%, and the aging of the camera pixels (BOL versus EOL) contributes less than 10\%.
\item
We wish to stress that using a prediction from planet formation models offers a unique possibility to validate these models against observations. The PLATO mission will deliver constraints to improve planet formation models.
\item
We find that a 3+1 scenario compared to a 2+2 scenario increases the PPY by roughly a factor of two
with a significant difference depending on orbital distance. To detect Earth-size planets with an orbit duration of about one year (our 250-500 d bin), the 2+2 case leads to a 50\% higher yield. 
\item
More planets can be detected around fainter stars of magnitude 12 or 13 compared to the brighter ones of magV 11 or brighter, which are more favorable to RV follow-up observations. The PPY ratios of these two magnitude ranges are 2.4 (NGPPS), 2.1 (HSU19), and 2.6 (KM20).
\item
We caution that the detection efficiency model may be subject to improvement in the course of mission preparation and after delivery of the first observational data from PLATO. Our procedure for computing the PPY can readily be adapted to other missions provided the detection efficiencies.
\end{enumerate}\vspace{0.5cm}
The final observing mission scenario will be determined two years before launch (see "red book"). This study provides a valuable assessment of PLATO's future mission performance and thus can facilitate this selection process.
\begin{acknowledgements}
      We warmly thank C. Mordasini for access to the NGPPS model prior to publication and for his comments and insight into the analysis which improved our scientific results.
      We thank G. Mulders for the provided support with the EPOS code, G.~Piotto for access to a preliminary PIC data base, and S.~Udry for useful discussions on the scientific results that improved their interpretation.
      We thank the 'transit' group at the EPA department of the Institute of Planetary Research at DLR for fruitful discussions.\\ 
      This work presents results from the European Space Agency (ESA) space mission PLATO. The PLATO payload, the PLATO Ground Segment and PLATO data processing are joint developments of ESA and the PLATO Mission Consortium (PMC). Funding for the PMC is provided at national levels, in particular by countries participating in the PLATO Multilateral Agreement (Austria, Belgium, Czech Republic, Denmark, France, Germany, Italy, Netherlands, Portugal, Spain, Sweden, Switzerland, Norway, and United Kingdom) and institutions from Brazil. Members of the PLATO Consortium can be found at https://platomission.com/. The ESA PLATO mission website is https://www.cosmos.esa.int/plato. We thank the teams working for PLATO for all their work. \\
      This research has made use of the NASA Exoplanet Archive, which is operated by the California Institute of Technology, under contract with the National Aeronautics and Space Administration under the Exoplanet Exploration Program.
\end{acknowledgements}

\bibliographystyle{aa} 
\bibliography{quellen} 

\begin{appendix}

\section{Average asPIC field versus LOPN1 and LOPS1}

We derived the number of stars in the pointing directions of the LOPN1 and LOPS1 fields (Nascimbeni et al. 2022) by fixing the rotation of the payload to the pointings ($Y_{PLM}$ pointing toward the north galactic pole). Then we determined whether the stars fall in the PLATO FoV.

In our derivation of the LOPN1 and LOPS1 fields, we obtain overall $+6,545$ (+4.1\%) and $+6,992$ (+4.5\%) more stars, respectively, than in the P5 sample according to Table 2 of Nascimbeni
et al. (2022). As Table A.1. shows, there are disproportionally more F stars and more faint stars in these pointing directions than on the average sky. 

\begin{table}[h!]
        \centering
        \caption{Number of stars in percent in the LOP1 fields}
        \begin{tabular}{ crrrr  }
                \hline
                magV & F & G & K \\
                \hline
                & N/S & N/S & N/S \\
                \hline
                8  &  +11.0/+14.4 & +1.6/+8.8 & +15.4/-6.0 \\
                9  &  +26.9/+22.3 & +5.9/+5.0 & +8.1/+14.4 \\
                10 &  +14.4/+30.2 & +13.5/+18.8 & +15.3/+5.3 \\
                11 &  +50.6/+35.9 & +27.7/+20.0 & +15.7/+12.4\\
                12 &  +43.8/+48.3 & +34.5/+22.7 & +19.7/+17.4 \\
                13 &  +37.5/+57.3 & +37.3/+25.0 & +25.4/+17.8 \\
                \hline
        \end{tabular}\\
        \label{tab:stars_in_lop1}
        \flushleft
        The number of stars is given in percentage deviation to 5.15\% of the respective number of stars in the asPIC according to our sorting for $n=0$; see Table \ref{tab:stars_in_pic}. $N$ denotes LOPN1, and $S$ denotes LOPS1.
\end{table}

With most of the PPY estimated to occur around magnitude 11-13 stars (Figure \ref{fig:ndet_magv}) and around F and G stars (Figure \ref{fig:HSU19_2_2_FGK}), the values in Table A.1. suggest a PPY enhancement by 30-40\% if these particular pointing directions were adopted.
\end{appendix}

\end{document}